\documentclass[conference]{IEEEtran}
\usepackage[boxruled]{algorithm2e}
\usepackage{cite}
\usepackage{graphicx}
\usepackage{psfrag}
\usepackage{subfigure}
\usepackage{url}
\usepackage{amsmath}
\usepackage{array}
\usepackage{amssymb}
\usepackage{amsfonts}
\usepackage{graphicx}
\usepackage{epstopdf}
\newtheorem{proposition}{Proposition}
\newtheorem{lemma}{Lemma}

\newtheorem{corollary}{Corollary}

\newtheorem{definition}{Definition}
\newtheorem{theorem}{Theorem}
\newtheorem{example}{Example}
\newtheorem{note}{Note}
\title{A Trellis Coded Modulation Scheme for the Fading Relay Channel}
\begin{document}

\author{
\authorblockN{Vijayvaradharaj T Muralidharan}
\authorblockA{Dept. of ECE, Indian Institute of Science \\
Bangalore 560012, India\\
Email: tmvijay@ece.iisc.ernet.in
}
\and
\authorblockN{B. Sundar Rajan}
\authorblockA{Dept. of ECE, Indian Institute of Science, \\Bangalore 560012, India\\
Email: bsrajan@ece.iisc.ernet.in
}
}

\maketitle
\thispagestyle{empty}	
\begin{abstract}
A decode and forward protocol based Trellis Coded Modulation (TCM) scheme for the half-duplex relay channel, in a Rayleigh fading environment, is presented. The proposed scheme can achieve any spectral efficiency  greater than or equal to one bit per channel use (bpcu).
A near-ML decoder for the suggested TCM scheme is proposed. It is shown that the high SNR performance of this near-ML decoder approaches the performance of the optimal ML decoder. The high SNR performance of this near-ML decoder is independent of the strength of the Source-Relay link and approaches the performance of the optimal ML decoder with an ideal Source-Relay link. Based on the derived Pair-wise Error Probability (PEP) bounds, design criteria to maximize the diversity and coding gains are obtained. 
Simulation results show a large gain in SNR for the proposed TCM scheme over uncoded communication as well as the direct transmission without the relay. Also, it is shown that even for the uncoded transmission scheme, the choice of the labelling scheme (mapping from bits to complex symbols) used at the source and the relay significantly impacts the BER vs SNR performance. We provide a good labelling scheme for $2^l$-PSK signal set, where $l\geq 2$ is an integer.
\end{abstract}
\section{PRELIMINARIES AND BACKGROUND}
\label{sec1}
\begin{figure}[htbp]
\centering
\includegraphics[totalheight=.75in,width=2in]{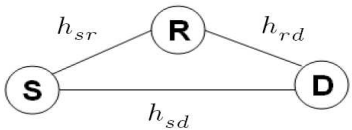}
\caption{The Relay Channel}	
\label{fig:Relay channel}	
\end{figure}

We consider the Rayleigh fading relay channel shown in Fig. \ref{fig:Relay channel}, consisting of the source node $S$, the relay node $R$ and the destination node $D$. It is assumed that R can operate only in the half-duplex mode, i.e., it cannot receive and transmit simultaneously in the same frequency band. It is assumed that R has perfect knowledge about the instantaneous value of the fade coefficient associated with the S-R link and D has perfect knowledge about the instantaneous values of the fade coefficients associated with the S-R, R-D and S-D links. Throughout, the phase during which the relay is in reception mode is referred to as Phase 1 and the phase during which the relay is in transmission mode is referred to as Phase 2.

 In all practical scenarios the S-R and R-D links are stronger than the S-D link. As a result, the use of coding schemes which involve both S and R, can potentially outperform the coding schemes involving S alone. The problem addressed in this paper is the design of a Trellis Coded Modulation (TCM) scheme for the relay channel, which achieves a spectral efficiency greater than or equal to one bit per channel use (bpcu) and provides a large gain compared to the TCM scheme for the direct transmission without the relay. 

A comparison of the proposed TCM scheme is made with the uncoded transmission scheme for the relay channel operating in the Decode and Forward (DF) mode, in which no coding is done at S and R, which is described as follows: At S, uncoded bits are directly mapped onto complex symbols from a signal set during Phase 1 and transmitted. During Phase 2, S and R map the uncoded bits (S uses the same bits as in Phase 1 and R uses the bits it decoded during Phase 1) onto complex symbols from a signal set for transmission. The signal sets used at S during Phase 1, S during Phase 2 and R during Phase 2 are assumed to be the same, but the mapping from bits to complex symbols can be different.
\subsection{The Proposed Scheme}
 \begin{figure*}[htbp]
\centering

\subfigure[Phase 1]{
\includegraphics[totalheight=1.5in,width=7in]{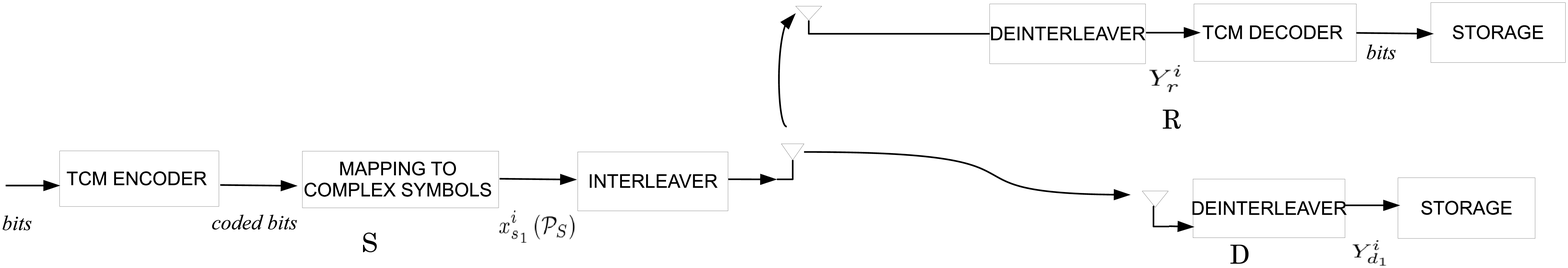}
\label{fig:phase1}	
}
\subfigure[Phase 2]{
\includegraphics[totalheight=1.5in,width=7in]{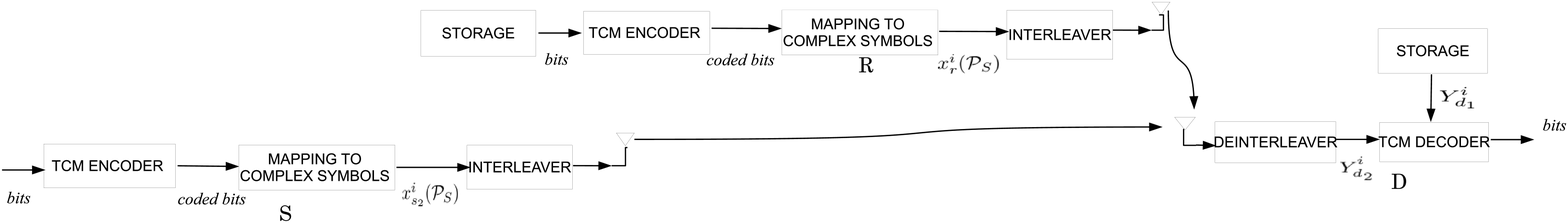}
\label{fig:phase2}	
}
\caption{The TCM Scheme}
\end{figure*}

The proposed TCM scheme for the fading relay channel is shown in Fig. \ref{fig:phase1} and Fig. \ref{fig:phase2} (shown at the top of the next page). It is assumed that TCM encoding at S during Phase 1 and Phase 2 and at R during Phase 2 take place using the same encoder, but the labelling schemes used for mapping coded bits onto signal points can be different. By a trellis, we refer to the states of the TCM encoder, the edges connecting these states in two successive stages and the complex numbers which are labelled on these edges. By a path in the trellis, we refer to the sequence of connected edges (the complex numbers labelled on the edges are not included). Let $\mathcal{T}_{S_1}$, $\mathcal{T}_{S_2}$ and $\mathcal{T}_{R}$ represent the trellises used for encoding by S during Phase 1, by S during Phase 2 and by R during Phase 2 respectively. Since the encoder corresponding to the trellises $\mathcal{T}_{S_1}$, $\mathcal{T}_{S_2}$ and $\mathcal{T}_{R}$ are the same, they differ only in the complex numbers labelled on the edges. Let ${\zeta}$ denote the set of edges connecting the states in two successive stages of the trellis. The labelling of the edges in the trellis $\mathcal{T}_{S_1}$ is given by the map $\mathcal{X}_{s_1}:\zeta \longrightarrow \mathcal{S}$, where $\mathcal{S}$ is the signal set used at S and R. Similarly, the maps $\mathcal{X}_{s_2}$ and $\mathcal{X}_{r}$ are defined for the the trellises $\mathcal{T}_{S_2}$ and $\mathcal{T}_{R}$ respectively.

During Phase 1, input bits at $S$ are encoded by the TCM encoder, mapped onto complex symbols from the signal set $\mathcal{S}$ and interleaved by the block interleaver as shown in Fig. \ref{fig:phase1}. Each transmission made is assumed to be a block of $L$ complex symbols. After each transmission, the encoder state is brought back to the all zeros state by appending zero input bits at the end. Let $\mathcal{P}_S$ denote the path in the trellis corresponding to the output of the TCM encoder at S. Let $ x_{s_1}^i(\mathcal{P}_S)$ denote the symbol to be transmitted corresponding to the $i^{th}$ branch of the path $\mathcal{P}_S$, where $1 \leq i \leq L$.

The received complex signal sequence at R and D during Phase 1 are given by,

\vspace{-0.25 cm}
{\footnotesize
\begin{align}
\left.
\begin{array}{ll}
\nonumber
Y_{r}^i&= h_{sr}^i x_{s_1}^i(\mathcal{P}_S)+z_{r};\\
\nonumber
Y_{d_1}^i&= h_{sd_1}^i x_{s_1}^i(\mathcal{P}_S)+z_{d_1};
\end {array}
\right\}
1 \leq i \leq L
\end{align}
}where  $h_{sr}^i$ and $h_{{sd}_1}^i$ are the independent zero mean circularly symmetric complex Gaussian fading coefficients associated with the S-R and S-D links respectively with the corresponding variances given by $\sigma^2_{sr}$ and $\sigma^2_{sd}$. The additive noises $z_{r}$ and $z_{d_1}$ are assumed to be $CN(0,1)$, where $CN(0,1)$ denotes the standard circularly symmetric complex Gaussian random variable. An interleaver with sufficiently large block length is assumed to make the block fading channel appear as a fading channel with independent fade coefficients for successive channel uses.  At R, after deinterleaving, an ML decoder using the Viterbi algorithm is used to decode the bits (Fig. \ref{fig:phase1}). Let $\mathcal{P}_R$ denote the path in the trellis to which R decodes. At D, the received complex numbers are deinterleaved and stored (Fig. \ref{fig:phase1}).

During Phase 2, S encodes the same bits used during Phase 1 and R encodes the bits it decoded during Phase 1, using the TCM encoder. The encoded bits at S and R are mapped onto complex symbols from the signal set $\mathcal{S}$ (possibly with different mappings and also possibly different from the mapping used at S during Phase 1) and interleaved using the block interleavers (same as the interleaver used at S during Phase 1) as shown in Fig. \ref{fig:phase2}. Let $x_{s_2}^i(\mathcal{P}_S)$ and $x_{r}^i(\mathcal{P}_R)$ denote the symbols transmitted by S and R respectively during Phase 2, corresponding to the $i^{th}$ branch of the paths $\mathcal{P}_S$ and $\mathcal{P}_R$, where $1 \leq i \leq L$.

The received complex signal sequence at D during Phase 2 is given by,

{\footnotesize
\begin {align}
\nonumber
Y_{d_2}^i&= h_{sd_2}^i x_{s_2}^i(\mathcal{P}_S)+h_{rd}^i x_{r}^i(\mathcal{P}_R)+z_{d_2},
\end {align}
}where $1 \leq i \leq L$. The random variables $h_{sd_2}^i$ and $h_{rd}^i$ are the independent zero mean circularly symmetric complex Gaussian fading coefficients associated with the S-D and R-D links respectively with the corresponding variances given by $\sigma^2_{sd}$ and $\sigma^2_{rd}$. The additive noise $z_{d_2}$ is $CN(0,1)$. Throughout, it is assumed that $\sigma^2_{sd}  \ll \sigma^2_{rd}$, due to the proximity of the relay to the destination than the source to the destination.

During Phase 2, at D the received complex numbers are deinterleaved and along with the complex numbers stored during Phase 1 are fed to the proposed near-ML TCM decoder described in Section III.

\subsection{Background and Related Work}
Achievable rates and upper bounds on the capacity for the discrete memoryless relay channel for the full-duplex and the more practical half-duplex relay channel were obtained in \cite{CoEl},\cite{KhSaAa1}. These results were extended for the half duplex Gaussian relay channel in \cite{KhSaAa2}. Optimal power allocation strategies to maximize the achievable rate for the Rayleigh fading relay channel were investigated in \cite{HoZh}. Several protocols like Amplify and Forward (AF), Decode and Forward (DF), Compress and Forward (CF) were proposed \cite{KrGaGu}. Power allocation strategies for the Non Orthogonal DF (NODF) protocol were discussed in \cite{YoYoYu}. Non-orthogonal relay protocols offer higher spectral efficiency when compared with orthogonal relay protocols \cite{NaBoKn}, \cite{AzGaSc}, \cite{ElViAnKu}. Hence, the proposed TCM scheme in this paper is based on the NODF protocol.

 A turbo-coding scheme for the  full duplex and more practical half-duplex relay channels were proposed in \cite{Turbo_full},\cite{Turbo_half}. These coding schemes achieve a spectral efficiency strictly less than one bpcu, i.e., they pay in bandwidth. Coding schemes for the relay channel with bandwidth constraints, which achieve a spectral efficiency greater than or equal to one bpcu are not reported in the literature, to the best of our knowledge. In this paper, we propose a Trellis Coded Modulation (TCM) scheme for the half-duplex relay channel for any spectral efficiency greater than or equal to one bpcu.
 
  Implementation as well as the performance analysis of the optimal ML decoder for the relay channel operating in the DF mode is complicated \cite{SeErAa},\cite{WaCaGiLa}. To solve these problems, several sub-optimal decoders for uncoded DF scheme were proposed in \cite{SeErAa}, \cite{WaCaGiLa}, \cite{ChLa} and \cite{JinJinNoShin}. These problems carry over to coded communication using DF protocol as well. If the decoder at D is designed assuming that R always decodes the bits correctly, the diversity order obtained is dependent on whether the relay decoded correctly or not. To circumvent this problem, relay selection strategies \cite{TaNo}, fountain codes \cite{XiTe}, Cyclic Redundancy Check (CRC) codes  \cite{JiAn} etc. are used. 
 To solve these problems of finding a decoder with implementable complexity and tractable performance analysis, we propose a near-ML decoder, whose performance at high SNR approaches the performance of the optimal ML decoder. The form of this decoder allows a DF scheme in which the relay need not check whether it has decoded correctly before forwarding the message.  This eliminates the need for embedding CRC bits as well as the need for feedback from the relay to the source. Furthermore, the diversity order of the proposed scheme does not depend on whether the relay decoded correctly or not. 

The proposed TCM scheme is different from the Co-operative Multiple Trellis Coded Modulation (CMTCM) scheme proposed in \cite{JiAn} in the following aspects:
\begin{itemize}
\item
The presence of embedded CRC bits and feedback from R to S is assumed in \cite{JiAn}, whereas we make no such assumptions.
\item
Interleaving/Deinterleaving is not assumed in \cite{JiAn} whereas we assume interleaving/deinterleaving which makes the quasi-static fading scenario appear as a fast fading scenario.
\end{itemize}

\subsection{Contributions and Organization}
The main contributions of this paper are as follows:
\begin{itemize}
\item
A TCM scheme for the fading relay channel is proposed which achieves any spectral efficiency greater than or equal to one bpcu. 
\item
A near-ML decoder is obtained for the proposed TCM scheme. The relay need not check whether it has decoded correctly before forwarding the message which eliminates the need for embedded CRC bits as well as the feedback from the relay to the source.
\item
The proposed near-ML decoder enables the formulation of design criteria to maximize the diversity order and the coding gain.
\item
Bounds on the PEP are derived for the near-ML decoder based on which the criteria to maximize the diversity order and coding gain are obtained. 
\item
The BER vs SNR performance of the near-ML decoder proposed with a non-ideal S-R link, at high SNR, approaches the performance of the optimal ML decoder with an ideal S-R link. This implies that the high SNR performance of the near-ML decoder for the relay channel with a non-ideal S-R link approaches the performance of the optimal ML decoder.
\item
It is shown that even for the uncoded transmission scheme, a proper choice of the labelling at S and R provides a significant performance improvement. We give a good labelling scheme for $2^l$-PSK signal set, where $l \geq 2$ is an integer.
\item
Simulation results show a large gain in the BER vs SNR performance for the proposed TCM scheme over the uncoded transmission scheme as well as the TCM scheme for the direct transmission without the relay. 
\end{itemize}

The organization of the rest of the paper is as follows: In Section II, the bounds on the ergodic capacity of the fading relay channel with Gaussian input alphabet are compared with the capacity bounds when the input is constrained to take values from a finite signal set. 
The description of the near-ML decoder for the TCM scheme constitutes Section III. In Section IV, the PEP expressions  are derived, based on which the code design criteria to maximize the diversity order and coding gain are obtained. In Section V, effect of the choice of the labelling scheme for the uncoded transmission scheme is discussed and a good labelling scheme for $2^l$-PSK constellation is presented. In Section VI, TCM code design examples are presented. Simulation results are presented in Section VII. 

\textit{\textbf{Notations:}} For a random variable $X_s$ which takes value from the set $\cal S$, we use $x_{s,i}$ to represent the $i$-th element of $\cal S$. $\textbf{E}_z[Y]$ denotes the expectation of $Y$ with respect to the random variable z. Throughout, $log$ refers to $log_2$ and $C(a)$ denotes $log(1+a)$. Let $CN(0,I_n)$ denote the standard circularly symmetric complex Gaussian random vector of length $n$. Also let $N(0,c)$ denote the scalar real valued Gaussian random variable with mean zero and variance $c$. For simplicity, distinction is not made between the random variable and a particular realization of the random variable, in expressions involving probabilities of random variables. 
In some probability expressions involving conditioning of the fading coefficients, the fact that the probability is conditioned on the values taken by the fading coefficients is not explicitly written, as it can be understood from the context. For a set $\mathcal{A}$, $\vert\mathcal{A}\vert$ denotes the cardinality of $\mathcal{A}$. For two sets $\mathcal{A}$ and $\mathcal{B}$, $\mathcal{A} \cup \mathcal{B}$ denotes the union of the sets $\mathcal{A}$ and $\mathcal{B}$. The superscript $(.)^T$ denotes the transpose operation. The set $\lbrace x \mid c(x) \rbrace$ denotes the set of all values of $x$ for which the condition $c(x)$ is satisfied. $\Re(x)$ denotes the real part of the complex number $x$. Throughout, $E_{\mathcal{S}}$ denotes the average energy in dB of the signal set $\mathcal{S}$ used at the source and the relay. 
Let Q[.] denote the complementary CDF of the standard Gaussian random variable.
\begin{figure*}
{\scriptsize
\begin{align}
\label{r4}
&R_1~\triangleq~I(X_{s_1};Y_r)]=log(M_{s_1})-\frac{1}{M_{s_1}}\sum_{i_1=0}^{M_{s_1}-1}\textbf{E}_{z_r}\left[log\left(\dfrac{\sum_{i = 0}^{M_{s_1}-1}exp\left(-\vert z_r-h_{sr}x_{s_1,i}+h_{sr}x_{s_1,i_1}\vert ^2\right)}{exp(-\vert z_r \vert^2)}\right)\right]\\
\label{r5}
&R_2~\triangleq~I(X_{s_2};Y_{d_2}|X_r)]=log(M_{s_2})-\frac{1}{M_{s_2}}\sum_{i_1=0}^{M_{s_2}-1}\textbf{E}_{z_{d_2}}\left[log\left(\dfrac{\sum_{i = 0}^{M_{s_2}-1}exp\left(-\vert z_{d_2}-h_{sd}x_{s_2,i}+h_{sd}x_{s_2,i_1}\vert ^2\right)}{exp(-\vert z_{d_2} \vert^2)}\right)\right]\\
\label{r6}
&R_3~\triangleq~I(X_{s_1};Y_{d_1})]=log(M_{s_1})-\frac{1}{M_{s_1}}\sum_{i_1=0}^{M_{s_1}-1}\textbf{E}_{z_{d_1}}\left[log\left(\dfrac{\sum_{i = 0}^{M_{s_1}-1}exp\left(-\vert z_{d_1}-h_{sd}x_{s_1,i}+h_{sd}x_{s_1,i_1}\vert ^2\right)}{exp(-\vert z_{d_1} \vert^2)}\right)\right]\\
\nonumber
\label{r7}
&R_4~\triangleq~I(X_{s_2},X_r;Y_{d_2})=log(M_{s_2}M_r)\\  
& \hspace{3 cm} -\dfrac{1}{M_{s_2}M_r}\sum_{i_1 =0}^{M_{s_2}-1}\sum_{j_1 = 0}^{M_r-1}\textbf{E}_{z_{d_2}}\left[log\left(\dfrac{\sum_{i = 0}^{M_{s_2}-1}\sum_{j =0}^{M_r-1}exp\left(-\vert z_{d_2}-h_{sd}x_{s_2,i}-h_{rd}x_{r,j}+h_{sd}x_{s_2,i_1}+h_{rd}x_{r,j_1}\vert ^2\right)}{exp(-\vert z_{d_2} \vert^2)}\right)\right]\\
\label{r8}
&R_5~\triangleq~I(X_{s_1};Y_r,Y_{d_1})]=log(M_{s_1})-\frac{1}{M_{s_1}}\sum_{i_1=0}^{M_{s_1}-1}\textbf{E}_{z_r ,z_{d_1}}\left[log\left(\dfrac{\sum_{i = 0}^{M_{s_1}-1}exp\left(-\vert z_r-h_{sr}x_{s_1,i}+h_{sr}x_{s_1,i_1}\vert ^2-\vert z_{d_1}-h_{sd}x_{s_1,i}+h_{sd}x_{s_1,i_1}\vert ^2\right)}{exp(-\vert z_r \vert^2-\vert z_{d_1} \vert^2)}\right)\right]\\
\hline
\nonumber
\end{align}}
\end{figure*}
\section{INFORMATION THEORETIC LIMITS} 

Throughout this section, the achievable rate of the decode and forward scheme is taken to be the lower bound on the capacity and the cut-set bound \cite{CoEl} is taken to be the upper bound on the capacity. 
%

For the half duplex relay channel, the received signal at $R$ during Phase 1 is given by,

{\footnotesize
\begin {align}
\nonumber
\hspace{-1.5 cm} Y_r  = h_{sr}X_{s_1} + z_r.
\end {align}
}
\noindent
The received signal at $D$ during Phase 1 and phase 2 are given by,

{\footnotesize
\begin {align}
\nonumber
&Y_{d_1} = h_{sd}X_{s_1} + z_{d_1};\\
\nonumber
&Y_{d_2} = h_{sd}X_{s_2} + h_{rd}X_r + z_{d_2}.
\end {align}
}where $X_{s_1}$ and $X_{s_2}$ denote the symbols transmitted by $S$ during Phase 1 and Phase 2 respectively, $X_r$ denotes the symbol transmitted by $R$ during Phase 2. The random variables $z_{d_1}$, $z_{d_2}$ and $z_r$ are independent and $CN(0,1)$.

For the relay channel with Gaussian input alphabet, the lower and upper bounds on the capacity are given by \cite{KhSaAa2} \cite{HoZh},

{\footnotesize
\begin{align}
\nonumber
C_{L}^G& =\dfrac{1}{2}\max_{0 \le \beta \le 1} \min \\
\nonumber
&\left\lbrace\textbf{E}_{h_{sr}}\left[\textbf{$C$}\left(\vert h_{sr}\vert ^2  E_\mathcal{S}\right)\right]
+\textbf{E}_{h_{sd}}\left[\textbf{$C$}\left((1-\beta)\vert h_{sd} \vert ^2 E_\mathcal{S}\right)\right],\right.\\
\nonumber
&\hspace{.2 cm}\textbf{E}_{h_{sd}}[\textbf{$C$}\left(\vert h_{sd}\vert ^2  E_\mathcal{S}\right)]\\
\label{ARG}
& \hspace{-30 cm} \left\lbrace \hspace{30cm}+\textbf{E}_{h_{sd},h_{rd}}\left[\textbf{$C$}\left(\vert h_{sd} \vert ^2 E_\mathcal{S}+{\vert h_{rd} \vert ^2 E_\mathcal{S}}+2\sqrt{\beta} \vert h_{sd}\vert \vert h_{rd} \vert E_\mathcal{S}\right)\right]
\right\rbrace;\\
\nonumber
C_{U}^G& = \dfrac{1}{2}\max_{0 \le \beta \le 1} \min
\left\lbrace
\textbf{E}_{h_{sd},h_{sr}}[\textbf{$C$}\left((\vert h_{sd}\vert ^2 +\vert h_{sr} \vert ^2) E_\mathcal{S}\right)] \hspace{100 cm}\right\rbrace\\
\nonumber
&\hspace{4 cm}+\textbf{E}_{h_{sd}}\left[\textbf{$C$}\left((1-\beta)\vert h_{sd} \vert ^2 E_\mathcal{S}\right)\right],\\
\nonumber
&\textbf{E}_{h_{sd},h_{rd}}\left[\textbf{$C$}\left(\vert h_{sd} \vert ^2  E_\mathcal{S}+ {\vert h_{rd} \vert ^2  E_\mathcal{S}}+2\sqrt{\beta} \vert h_{sd}\vert \vert h_{rd}\vert E_\mathcal{S}\right)\right]\\
\label{UBG}
&\hspace{-20 cm} \left\lbrace\hspace{24.5 cm}+\textbf{E}_{h_{sd}}\left[\textbf{$C$}\left(\vert h_{sd}\vert ^2  E_\mathcal{S}\right)\right]
\right\rbrace.
\end{align}
}


To compute the achievable rate bounds with finite input constellation, assume  $X_{s_1} \in \mathcal{S}_{s_1}$, $ X_{s_2} \in \mathcal{S}_{s_2}$ and $X_{r} \in \mathcal{S}_{r}$, where $\mathcal{S}_{s_1}$, $\mathcal{S}_{s_2}$ and $\mathcal{S}_{r}$ are signal sets such that $\vert \mathcal{S}_{s_1}\vert =M_{s_1}$, $\vert\mathcal{S}_{s_2}\vert = M_{s_2}$ and $\vert\mathcal{S}_{r}\vert= M_{r}$.
Define $R_4$ = $I(X_{s_1};Y_r)$, $R_5$ = $I(X_{s_2};Y_{d_2}|X_r)$, $R_6$ = $I(X_{s_1};Y_{d_1})$, $R_7$ = $I(X_{s_2},X_r;Y_{d_2})$ and $R_8$ = $I(X_{s_1};Y_r,Y_{d_1})$. It is assumed that coding schemes in which all the possible choices for $(X_{s_1},X_{s_2},X_r)$ occur with equal probability only are of interest. This forces the marginal distributions of $X_{s_1}$, $X_{s_2}$, $X_r$ to be independent and uniformly distributed. 

 
 The achievable rate bounds of the half duplex relay channel with finite input constellation can be shown to be \cite{VvR_arxiv},

{\footnotesize
\begin{align}
\nonumber
&C_{L}^{CC} = \min
\left\lbrace
\dfrac{1}{2}\textbf{E}_{h_{sr}}\left[R_1\right]+\dfrac{1}{2}\textbf{E}_{h_{sd}}\left[R_2\right],\hspace{100 cm}\right\rbrace\\
\label{ARHD}
&\hspace{-25 cm}\left\lbrace\hspace{28.5 cm} \dfrac{1}{2}\textbf{E}_{h_{sd}}\left[R_3\right]+\dfrac{1}
{2}\textbf{E}_{h_{sd},h_{rd}}\left[R_4\right]
\right\rbrace;\\
\nonumber
&C_{U}^{CC} = \min
\left\lbrace
\dfrac{1}{2}\textbf{E}_{h_{sr},h_{sd}}\left[R_5\right]+\dfrac{1}{2}\textbf{E}_{h_{sd}}\left[R_2\right],\hspace{100 cm}\right\rbrace\\
\label{UBHD}
&\hspace{-25 cm} \left\lbrace\hspace{28.5 cm}\dfrac{1}{2}\textbf{E}_{h_{sd}}\left[R_3\right]+\dfrac{1}{2}\textbf{E}_{h_{sd},h_{rd}}\left[R_4\right]
\right\rbrace.
\end{align}
}where $R_1$-$R_5$ are given by \eqref{r4}-\eqref{r8}, shown at the top of this page.

\begin{figure}[htbp]
\vspace{-0.3 cm}
\centering
\includegraphics[totalheight=3in,width=4in]{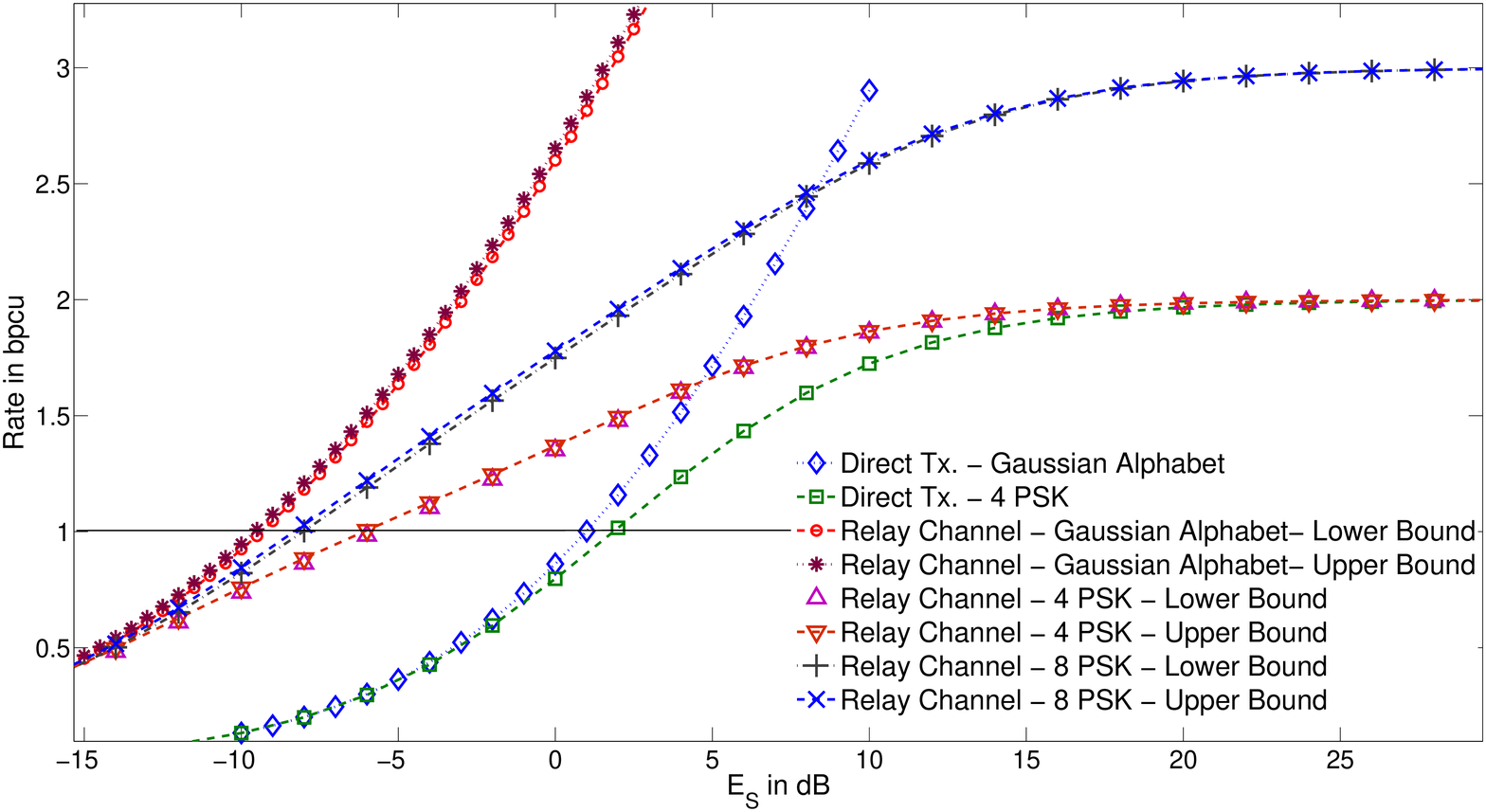}
\vspace{-0.7 cm}
\caption{Rate vs $E_\mathcal{S}$ for Half Duplex relay Channel with $\sigma^2_{sd}$=0 dB, $\sigma^2_{sr}$=15 dB, $\sigma^2_{rd}$=15 dB}	
\label{fig:TD_capacity}	
\vspace{-0.4 cm}
\end{figure}

The expressions for the capacity bounds \eqref{ARG}, \eqref{UBG}, \eqref{ARHD} and \eqref{UBHD} can be evaluated using Monte-Carlo simulations. The plots for the $E_\mathcal{S}$ Vs Capacity bounds for the relay channel with Gaussian alphabet as well as 4-PSK and 8-PSK constellations, for $\sigma^2_{sd}$=0 dB, $\sigma^2_{sr}$=15 dB, $\sigma^2_{rd}$=15 dB, are shown in Fig. \ref{fig:TD_capacity}. Fig. \ref{fig:TD_capacity} also shows the capacity with Gaussian alphabet and the constellation constrained capacity with 4-PSK signal set for the direct transmission (S-D link). From Fig. \ref{fig:TD_capacity}, it is clear that the use of relay can provide significant advantage over the direct transmission. From Fig. \ref{fig:TD_capacity}, we can see that  for a spectral efficiency of 1 bpcu, for $\sigma^2_{sd}$=0 dB, $\sigma^2_{sr}$=15 dB, $\sigma^2_{rd}$=15 dB, the achievable rate using DF scheme for 4-PSK and 8-PSK input constellations are about 3.5 dB and 1.5 dB away from that of the Gaussian alphabet. Hence in order to be within 1.5 dB away from the achievable rate of the DF scheme using Gaussian alphabet, for a spectral efficiency equal to 1 bpcu, 8 point signal set needs to be used. Hence, in the proposed TCM scheme, to achieve a spectral efficiency $r$ bpcu, a signal set with $2^{2r+1}$ points is used.
\begin{note}
To achieve a spectral efficiency of $r$ bpcu, a total of $2r$ bits get transmitted from S to D in the two Phases of relaying.
\end{note}

The value of ${E}_{\mathcal{S}}$ required for a BER of $10^{-4}$ and the diversity order for the different schemes achieving a spectral efficiency of 1 bpcu are shown in Table \ref{table5}, at the top of the next page (The details regarding the design of TCM schemes for the relay channel are presented in Section VI and $E_\mathcal{S}$ vs $BER$ plots are given in Section VII. The TCM schemes for the direct transmission were 
designed based on the design criteria for TCM for fading channels \cite{Da}). From Table \ref{table5}, it can be seen that the proposed schemes for the relay channel, which exploit the fact that the S-R and R-D links are better than the S-D link, outperform the corresponding schemes for the direct transmission without relay. From Fig. \ref{fig:TD_capacity}, we see that  the capacity of the direct transmission without relay with 4 PSK signal set and Gaussian alphabet are respectively 2 dB and 1.5 dB, which are greater than the value of $E_\mathcal{S}$ (0 dB) required for the 16 state TCM scheme for the relay channel, for a BER of $10^{-4}$. This means that under the given assumptions, the proposed 16 state 8 PSK TCM scheme for the relay channel, will outperform even the best possible coding scheme for the direct transmission without relay. The claim made above is valid even for other values of the variances of the fading links, possibly with a higher number of states if not with 16 states. 
\begin{table*}
\centering
  \caption{Comparison of different schemes for a spectral efficiency of 1 bpcu }
 \label{table5}
 \begin{tabular}{|c|c|c|c|}

\hline Scheme & Diversity Order & $E_\mathcal{S}$ in dB for $BER=10^{-4}$\\ 
\hline Direct Tx. - BPSK                                           & 1 & 37 dB\\ 
\hline Direct Tx. - 2 State TCM                                    & 2 & 18 dB\\ 
\hline Direct Tx. - 4 State TCM                                    & 3 & 12 dB\\
\hline Relay Channel - Uncoded Tx. Scheme                          & 2 & 12.5 dB\\
\hline Relay Channel - 2 State TCM                                 & 2 & 8  dB\\
\hline Relay Channel - 4 State TCM                                 & 4 & 3  dB\\
\hline Relay Channel - 8 State TCM                                 & 4 & 1.5 dB\\
\hline Relay Channel - 16 State TCM                                & 6 & 0 dB\\
\hline 
\end{tabular} 
\end{table*}

\section{A NEAR-ML TCM DECODER}

\begin{figure}[htbp]
\centering
\includegraphics[totalheight=2in,width=2in]{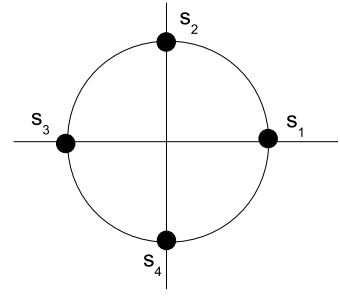}
\caption{4-PSK signal set}	
\label{fig:4psk}	
\end{figure}
\begin{figure}[htbp]
\centering
\includegraphics[totalheight=3in,width=3in]{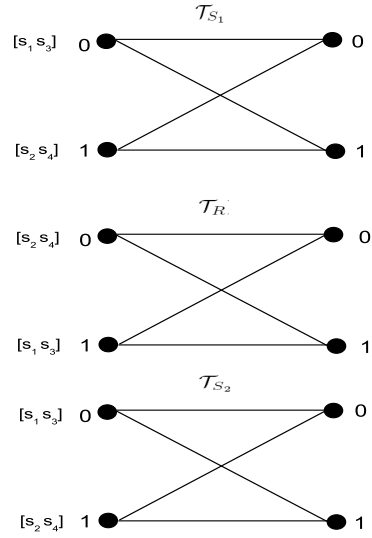}
\caption{Trellises $\mathcal{T}_{S_1}$, $\mathcal{T}_{R}$ and $\mathcal{T}_{S_2}$}	
\label{fig:trellis_example}	
\end{figure}
\begin{figure}[htbp]
\centering
\includegraphics[totalheight=3in,width=4in]{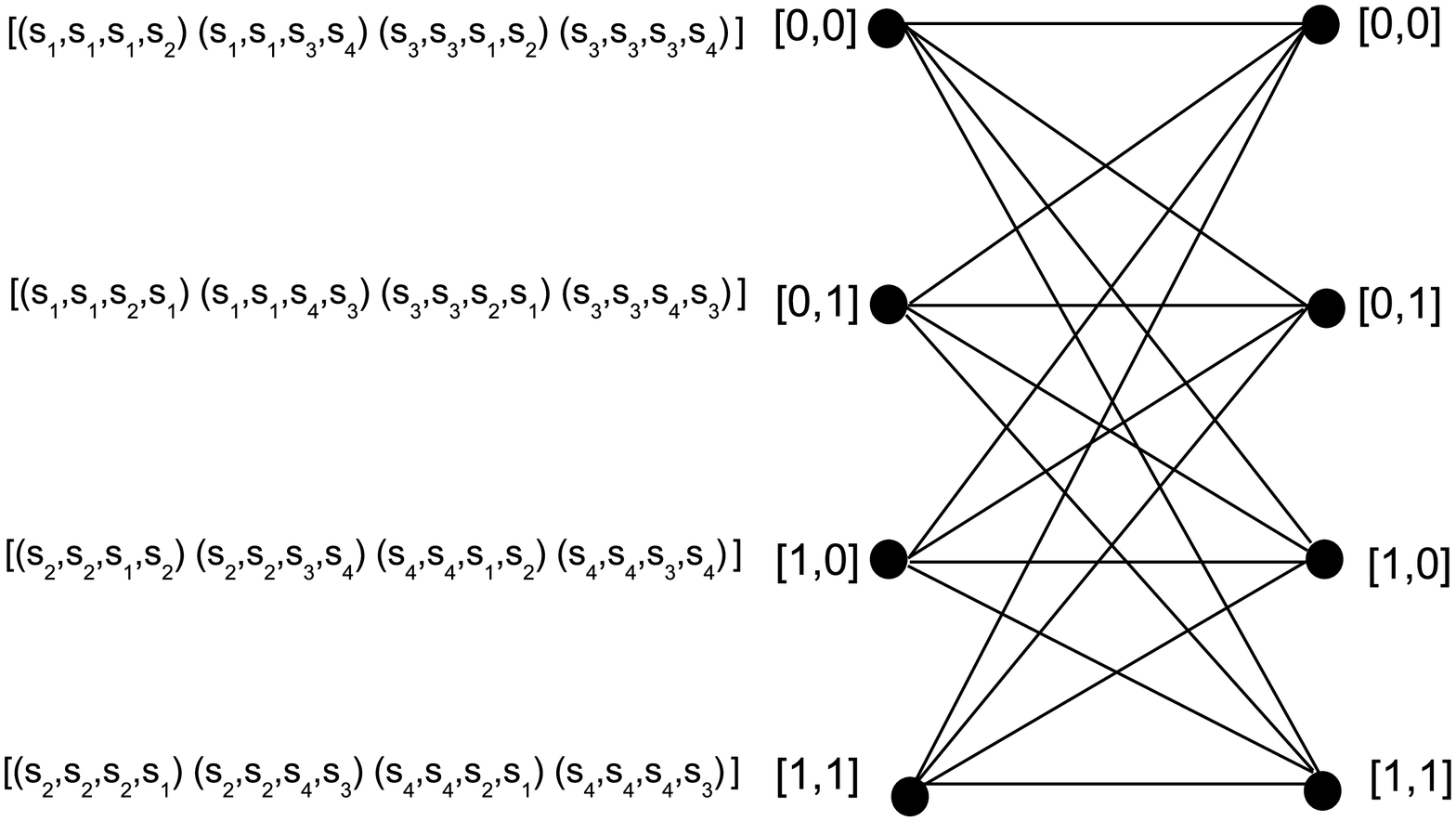}
\caption{Trellis $\mathcal{T}_{D}$ used for decoding at D}	
\label{fig:trellis_decoder}	
\end{figure}

The trellis used for decoding at D is constructed as described in the following subsection. The reason why the decoding trellis is of the form described will be clear by the end of Subsection B of this section.
\subsection{Decoding Trellis at D}
  Let $\lbrace a^i,a^{i+1}\rbrace$ and $\lbrace b^i,b^{i+1}\rbrace$ respectively denote any two pairs of states in the $i^{th}$ and $i+1^{th}$ stages of the trellises $\mathcal{T}_{S_1}$ and $\mathcal{T}_{R}$. Let $\mathcal{E}_{ a^i,a^{i+1}}$ and $\mathcal{E}_{ b^i,b^{i+1}}$ denote the sets of edges from $a^i$ to $a^{i+1}$ and $b^i$ to $b^{i+1}$ respectively. 

 The trellis used for decoding at D, denoted as $\mathcal{T}_D$, is constructed as follows: The states of $\mathcal{T}_D$ at the $i^{th}$ and $i+1^{th}$ stages are denoted by the two tuples $[a^i,b^i]$ and $[a^{i+1},b^{i+1}]$ respectively. The set of edges from  $[a^i,b^i]$ to $[a^{i+1},b^{i+1}]$, denoted as $\mathcal{E}_{[a^i,b^i],[a^{i+1},b^{i+1}]}$, contains an edge denoted by the pair $\lbrace e_{ a^i,a^{i+1}},e_{b^i,b^{i+1}}\rbrace$ if $e_{ a^i,a^{i+1}} \in  \mathcal{E}_{ a^i,a^{i+1}}$ and $e_{ b^i,b^{i+1}} \in  \mathcal{E}_{ b^i,b^{i+1}}$. The edge $\lbrace e_{ a^i,a^{i+1}},e_{ b^i,b^{i+1}}\rbrace$ is labelled with the four tuple $(\mathcal{X}_{s_1}(e_{ a^i,a^{i+1}}),\mathcal{X}_{s_2}(e_{ a^i,a^{i+1}}),\mathcal{X}_{s_1}(e_{ b^i,b^{i+1}}),\mathcal{X}_{r}(e_{ b^i,b^{i+1}})) \in \mathcal{S}^4$. 
\begin{note}
Even though encoding at S and R during the two Phases take place using a total of three trellises, the edges connecting the states in two successive stages in the trellis $\mathcal{T}_D$ are constructed based on the trellises $\mathcal{T}_{S_1}$ and $\mathcal{T}_{R}$. The reason is as follows: The uncoded bits at the input of the TCM encoders at S during Phase 1 and Phase 2 are the same. Hence, the encoded paths in the trellises $\mathcal{T}_{S_1}$ and $\mathcal{T}_{S_2}$ are always the same (though the complex signal sequences may be different, due to different labellings) and it is enough to consider one of the two trellises $\mathcal{T}_{S_1}$ and $\mathcal{T}_{S_2}$.
\end{note}

\begin{example} 
For the trellis triple $\lbrace \mathcal{T}_{S_1}, \mathcal{T}_{S_2}, \mathcal{T}_{R} \rbrace$ (shown in Fig. \ref{fig:trellis_example}), labelled with signal points from $\mathcal{S}$ (shown in Fig. \ref{fig:4psk}), the trellis $\mathcal{T}_D$  is shown in Fig. \ref{fig:trellis_decoder}.
\end{example}
%

The decoding metric for the proposed near-ML decoder is obtained in the following subsection. 
\subsection{Near-ML Decoding Metric} 
By assumption, R has perfect knowledge about the instantaneous
value of the fade coefficient associated with the  S-R link
and D has perfect knowledge about the instantaneous values
of the fade coefficients associated with the S-R, R-D and S-D
links. Let $\mathcal{A}_L$ denote the set of all paths of length $L$ in the trellis (since the TCM encoder used at S during Phase 1, at S during Phase 2 and at R during Phase 2 are the same, the set $\mathcal{A}_L$ is the same for the trellises $\mathcal{T}_{S_1}$, $\mathcal{T}_{S_2}$ and $\mathcal{T}_{R}$).
 Let $\mathcal{P}_S$ denote the path which corresponds to the symbols transmitted by the source during Phase 1. Let $\mathcal{P}_R$ denote the path to which the Viterbi decoder at $R$ decodes during Phase 1, i.e.,
 
{\footnotesize 
 \begin{align}
 \nonumber
\mathcal{P}_R= \arg\min_{\mathcal{P}\in \mathcal{A}_L} \sum_{i=1}^{L}\left\vert Y_{r}^i-h_{sr}^ix_{s_1}^i\left(\mathcal{P}\right)\right\vert^2.
\end{align}
}
 
 The optimal ML decoder at D decides in favour of the path,
 
 {\footnotesize
 \begin{align}
 \nonumber
 \mathcal{\hat{P}_S}=\arg \max_{\mathcal{P}_S\in \mathcal{A}_L} Pr\left\lbrace Y_{d_1}^{1,L},Y_{d_2}^{1,L}\vert\mathcal{P}_S\right\rbrace ,
 \end{align}
 }where $Y_{d_1}^{1,L}$ and $Y_{d_2}^{1,L}$ denote the sequences $Y_{d_1}^1, Y_{d_1}^2,....,Y_{d_1}^L$ and $Y_{d_2}^1, Y_{d_2}^2,....,Y_{d_2}^L$ respectively and $Pr\left\lbrace Y_{d_1}^{1,L},Y_{d_2}^{1,L}\vert\mathcal{P}_S\right\rbrace$ is the probability that $Y_{d_1}^{1,L}$ and $Y_{d_2}^{1,L}$ are the received sequences by D during Phase 1 and Phase 2, given that the path corresponding to the complex numbers transmitted by S is $\mathcal{P}_S$. The probability which needs to be maximized is given by,
 
 {\footnotesize
 \begin{align}
 \label{eqn1}
 Pr\left\lbrace Y_{d_1}^{1,L},Y_{d_2}^{1,L}\vert\mathcal{P}_S\right\rbrace = \sum_{\mathcal{P}_R\in \mathcal{A}_L}Pr\left\lbrace Y_{d_1}^{1,L},Y_{d_2}^{1,L}\vert{\mathcal{P}_S},{\mathcal{P}_R}\right\rbrace Pr\left\lbrace{\mathcal{P}_R}\vert{\mathcal{P}_S}\right\rbrace.
 \end{align}
 }where $Pr\left\lbrace Y_{d_1}^{1,L},Y_{d_2}^{1,L}\vert{\mathcal{P}_S},{\mathcal{P}_R}\right\rbrace$ is the probability that $Y_{d_1}^{1,L}$ and $Y_{d_2}^{1,L}$ are the received sequences by D during Phase 1 and Phase 2, given that the paths corresponding to the complex numbers transmitted by S and R are $\mathcal{P}_S$ and $\mathcal{P}_R$ respectively. The quantity $Pr\left\lbrace{\mathcal{P}_R}\vert{\mathcal{P}_S}\right\rbrace$ is the probability that R decodes to the path $\mathcal{P}_R$ given that $\mathcal{P}_S$ is the path corresponding to the complex symbols transmitted by S. 
 
 The probability $Pr\left\lbrace \mathcal{P}_R\vert\mathcal{P}_S\right\rbrace$ can be upper bounded by the corresponding PEP, i.e., 
 
 {\footnotesize
\begin{align} 
 \nonumber
 Pr\left\lbrace{\mathcal{P}_R}\vert{\mathcal{P}_S}\right\rbrace &\leq \prod_{i=1}^{L}Q\left[\dfrac{\left\vert h_{sr}^i \left(x_{s_1}^i\left(\mathcal{P}_R\right)-x_{s_1}^i\left(\mathcal{P}_S\right)\right)\right\vert} {\sqrt{2}} \right ]\\
 \label{eqn2}
 &\leq \exp\left\lbrace-\dfrac{1}{4}\sum_{i=1}^{L}\left\vert h_{sr}^i \left(x_{s_1}^i\left(\mathcal{P}_S\right)-x_{s_1}^i\left(\mathcal{P}_R\right)\right)\right\vert^2\right\rbrace,
 \end{align}
 }and we also have,
 
 {\footnotesize
 \begin{align}
 \nonumber
 Pr\left\lbrace Y_{d_1}^{1,L},Y_{d_2}^{1,L}\vert{\mathcal{P}_S},{\mathcal{P}_R}\right\rbrace &=\\
\nonumber 
  &\hspace{-2 cm}\dfrac{1}{\pi^{2L}}\exp\left\lbrace-\sum_{i=1}^{L}\left(\left\vert Y_{d_1}^i-h_{sd_1}^i x_{s_1}^i\left(\mathcal{P}_S\right)\right\vert^2 \hspace{100 cm}\right)\right\rbrace\\
  \label{eqn3}
&\hspace{-32 cm}\left\lbrace\left(\hspace{31 cm}+\left\vert Y_{d_2}^i-h_{sd_2}^i x_{s_2}^i\left(\mathcal{P}_S\right)-h_{rd}^i x_{r}^i\left(\mathcal{P}_R\right)\right\vert^2\right)\right\rbrace.
   \end{align}
 }
 
 Substituting \eqref{eqn2} and \eqref{eqn3} in \eqref{eqn1} we get \eqref{eqn4}, shown at the top of the next page.
 
 \begin{figure*}
{\scriptsize
\begin{align}
\label{eqn4}
&Pr\left\lbrace Y_{d_1}^{1,L},Y_{d_2}^{1,L}\vert\mathcal{P}_S\right\rbrace \leq\dfrac{1}{\pi^{2L}}\sum_{\mathcal{P}_R\in \mathcal{A}_L}\exp\left\lbrace-\sum_{i=1}^{L}\left(\left\vert Y_{d_1}^i-h_{sd_1}^i x_{s_1}^i\left(\mathcal{P}_S\right)\right\vert^2 +\left\vert Y_{d_2}^i-h_{sd_2}^i x_{s_2}^i\left(\mathcal{P}_S\right)-h_{rd}^i  x_{r}^i\left(\mathcal{P}_R\right)\right\vert^2+\dfrac{1}{4}\left\vert h_{sr}^i \left(x_{s_1}^i\left(\mathcal{P}_R\right)-x_{s_1}^i\left(\mathcal{P}_S\right)\right)\right\vert^2\right)\right\rbrace\\
\hline
 \label{eqnphi}
&\phi_i\left(\mathcal{P}_S,\mathcal{P}_R\right)=\left\vert Y_{d_1}^i-h_{sd_1}^i x_{s_1}^i\left(\mathcal{P}_S\right)\right\vert^2 +\left\vert Y_{d_2}^i-h_{sd_2}^i x_{s_2}^i\left(\mathcal{P}_S\right)-h_{rd}^i x_{r}^i\left(\mathcal{P}_R\right)\right\vert^2+\dfrac{1}{4}\left\vert h_{sr}^i \left(x_{s_1}^i\left(\mathcal{P}_S\right)-x_{s_1}^i\left(\mathcal{P}_R\right)\right)\right\vert^2
\end{align}}
\hrule
\end{figure*} 

If we maximize only the dominant exponential in the upper bound \eqref{eqn4}, then the decoded path is given by,

 {\footnotesize
 \begin{align}
 \label{eqn5}
\mathcal{\hat{P}}_S=\arg \min_{\mathcal{P}_S\in \mathcal{A}_L}\left\lbrace\min_{\mathcal{P}_R\in \mathcal{A}_L}\sum_{i=1}^L\phi_i\left(\mathcal{P}_S,\mathcal{P}_R\right)\right\rbrace,
\end{align}
}where the metric $\phi_i\left(\mathcal{P}_S,\mathcal{P}_R\right)$ is given by \eqref{eqnphi} (shown at the top of the next page).

This near-ML decoder given by \eqref{eqn5} involves minimizing the additive metric over two distinct paths and hence the decoding takes place in the trellis $\mathcal{T}_D$ described in the previous subsection.
%
Throughout, it is assumed that decoding at D takes place using this near-ML decoder.

The branch metric for the Viterbi decoder corresponding to the edge $\lbrace e_{ a^i,a^{i+1}},e_{ b^i,b^{i+1}}\rbrace$ in $\mathcal{T}_D$ can be obtained from \eqref{eqnphi} and is given by 

{\footnotesize
\begin{align}
\nonumber
&f^i\left( e_{ a^i,a^{i+1}},e_{ b^i,b^{i+1}}\right)=\left\vert Y_{d_1}^i-h_{sd_1}^i \mathcal{X}_{s_1}(e_{ a^i,a^{i+1}})\right\vert^2 \\
\nonumber
&\hspace{2.2 cm}+\left\vert Y_{d_2}^i-h_{sd_2}^i \mathcal{X}_{s_2}(e_{ a^i,a^{i+1}})-h_{rd}^i \mathcal{X}_{r}(e_{ b^i,b^{i+1}})\right\vert^2\\
\nonumber
&\hspace{3 cm}+\dfrac{1}{4}\left\vert h_{sr}^i \left(\mathcal{X}_{s_1}(e_{ a^i,a^{i+1}})-\mathcal{X}_{s_1}(e_{ b^i,b^{i+1}})\right)\right\vert^2.
\end{align}
}

For example, for the edge from state $[1,0]$ to state $[0,1]$ in $\mathcal{T}_D$ in Example 1, denoted as $\lbrace e_{1,0},e_{0,1}\rbrace$, from Fig. \ref{fig:trellis_decoder}, the four tuple $(\mathcal{X}_{s_1}(e_{ 1,0}),\mathcal{X}_{s_2}(e_{1,0}),\mathcal{X}_{s_1}(e_{0,1}),\mathcal{X}_{r}(e_{0,1}))$ is $(s_2,s_2,s_3,s_4)$ and hence the decoding metric is given by,

{\footnotesize
\begin{align}
\nonumber
&f^i\left( e_{1,0},e_{0,1}\right)=\left\vert Y_{d_1}^i-h_{sd_1}^i s_2\right\vert^2+\left\vert Y_{d_2}^i-h_{sd_2}^i s_2-h_{rd}^i s_4\right\vert^2\\
\nonumber
&\hspace{5.8 cm}+\dfrac{1}{4}\left\vert h_{sr}^i \left(s_2-s_3\right)\right\vert^2.
\end{align}
}

Let $K$ denote the number of branches which originate from each state and let $N$ denote the number of states in the trellises $\mathcal{T}_{S_1}$, $\mathcal{T}_{S_2}$ and $\mathcal{T}_{R}$,. In the $N^2$ state trellis $\mathcal{T}_D$, $K^2$ branches originate from a state. As a result, in each step, the Viterbi algorithm used for decoding involves $N^2K^2$ additions and $N^2$ comparisons of $K^2$ values. Hence the decoding complexity of the proposed near-ML decoder is approximately $2N^2K^2/(log_2K)$ operations per decoded bit, since $\log_2K$ bits get transmitted in the two phases of relaying. 
\section{PEP ANALYSIS OF THE NEAR-ML DECODER AND CODE DESIGN CRITERIA}

The following Lemma is useful for the PEP analysis of the near-ML decoder.
\begin{lemma}
\label{lemma1}
Consider a communication system with one transmitter and one receiver. The transmitter transmits one of the two  $n$-dimensional complex vectors  $x_1$ and $x_2$ $\in$ $\mathbb{C}^n$ corresponding to two messages. The received $n$-dimensional complex vector $y=x+z$ $\in$ $\mathbb{C}^n$, where $x$ $\in$ $\lbrace x_1,x_2 \rbrace$  and $z$ is $CN(0,I_n)$. The decoding rule used at the receiver is as follows: The decoder decides in favour of message $1$,
if $||y-x_1||^2 \leq ||y-x_2||^2 + c$
and it decides in favour of message $2$ otherwise,
where $c \in \mathbb{R}$ is a constant. Then the probability that the decoder decides in favour of message $2$, given that message $1$ was transmitted is upper bounded as,

{
\begin{align}
\nonumber
Pr \left(1 \longrightarrow 2 \right) \leq \dfrac{1}{2} \exp \left \lbrace -\dfrac{||x_1-x_2||^2}{4}-\dfrac{c}{2}\right\rbrace.
\end {align}
}

\begin{proof}

We have,

{\footnotesize
\begin{align}
\nonumber
Pr \left(1 \longrightarrow 2 \right)&=Pr\left(\left(||y-x_1||^2 \geq ||y-x_2||^2 + c \right)\mid x=x_1 \right)\\
\nonumber
&=Pr\left(\Re\left \lbrace \left (y- \dfrac {x_1+x_2} {2}\right)^* \left(x_2 - x_1\right)  \right\rbrace \geq \dfrac{c}{2}\right\vert x=x_1)\\
\nonumber
&=Pr\left(\Re\left \lbrace \left (z+\dfrac {x_1-x_2} {2}\right)^* \left(x_2 - x_1\right)  \right\rbrace \geq \dfrac{c}{2}\right)\\
\nonumber
&=Pr\left(\Re\left \lbrace z^* \dfrac {x_2-x_1}{\vert\vert x_2 - x_1 \vert\vert} \right\rbrace\hspace{30 cm}\right) \\
\nonumber
&\hspace{-15 cm}\left(\hspace{17.4 cm}\geq \dfrac{c}{2{\vert\vert x_2 - x_1 \vert\vert}}+{\dfrac{\vert\vert x_2 - x_1 \vert\vert}{2}}\right)
\end{align}
}
Since $z$ is $CN$(0,$I_n$), it can be shown that $\Re\left\lbrace z^*\dfrac {x_2-x_1}{\vert\vert x_2 - x_1 \vert\vert}\right\rbrace$ is  $N\left(0,\dfrac{1}{2}\right)$.
\newline
Hence,

{\footnotesize
\begin{align}
\nonumber
Pr \left(1 \longrightarrow 2 \right)&=Q\left[\sqrt{2}\left(\dfrac{\vert\vert x_1-x_2 \vert \vert}{2}+\dfrac{c}{2{\vert\vert x_2 - x_1 \vert\vert}}\right)\right]\\
\nonumber
&\leq\exp\left\lbrace-\left(\dfrac{\vert\vert x_1-x_2 \vert \vert}{2}+\dfrac{c}{2{\vert\vert x_2 - x_1 \vert\vert}}\right)^2\right\rbrace\\
\nonumber
&\leq \exp \left \lbrace -\dfrac{||x_1-x_2||^2}{4}-\dfrac{c}{2}\right\rbrace.
\end{align}
}
\end{proof}
\end{lemma}
\subsection{PEP Analysis of the Near-ML Decoder}
In this subsection, an upper bound on the PEP that a transmitted path $\mathcal{P}_S$ at S is decoded by the near-ML decoder at D as ${\mathcal{\tilde{P}}_{S}}$ is derived.

For the trellis $\mathcal{T}_{S_1}$, let $\eta_{s_1}(\mathcal{P}_S,\mathcal{\tilde{P}}_S)$ denote the set of values of $i$ for which $x_{s_1}^i(\mathcal{P}_S)$ and $x_{s_1}^i(\mathcal{\tilde{P}}_S)$ are different, for $1 \leq i \leq L$. Similarly,  $\eta_{s_2}(\mathcal{P}_S,\mathcal{\tilde{P}}_S)$ and $\eta_{r}(\mathcal{P}_S,\mathcal{\tilde{P}}_S)$ are defined for the trellises $\mathcal{T}_{S_2}$ and $\mathcal{T}_{R}$ respectively.

\begin{theorem}
For the proposed TCM scheme, the PEP that a transmitted path $\mathcal{P}_S$ in the trellis is decoded as $\mathcal{\tilde{P}}_S$, under the proposed near-ML decoder, is upper bounded by \eqref{eqn11_mod} (shown at the top of the next page).

\begin{figure*}
\scriptsize
\begin{align}
\nonumber
&Pr\left\lbrace\mathcal{P}_S\longrightarrow{\mathcal{\tilde{P}}_{S}}\right\rbrace  \leq  \left(\prod_{i \in \eta_{s_1}
(\mathcal{P}_S,\mathcal{\tilde{P}}_S)} \left[\dfrac {1}{1+\dfrac{\left\vert \sigma_{sd}\left(x_{s_1}^i\left(\mathcal{P}_S\right)-x_{s_1}^i(\mathcal{\tilde{P}}_S)\right)\right\vert^2}{4}}\right]\right.\\
\nonumber
&\hspace{7 cm} \left.\prod_{i \in \left \lbrace \eta_{s_2}(\mathcal{P}_S,\mathcal{\tilde{P}}_S) \cup \eta_{r}(\mathcal{P}_S,\mathcal{\tilde{P}}_S)\right \rbrace}\left[\dfrac{1}{1+\dfrac{\left\vert \sigma_{sd}\left(x_{s_2}^i\left(\mathcal{P}_S\right)-x_{s_2}^i(\mathcal{\tilde{P}}_S)\right)\right\vert^2+\left\vert\sigma_{rd}\left(x_{r}^i\left(\mathcal{P}_S\right)-{x}_{r}^{i}(\mathcal{\tilde{P}}_S)\right)\right\vert^2}{4}}\right]\right)\\
 \nonumber
 &+\sum_{\substack{{\mathcal{{P}}_R\in \mathcal{A}_L} \\ { \mathcal{{P}}_R \neq \mathcal{{P}}_S}}}\sum_{\substack{{\mathcal{\tilde{P}}_R\in \mathcal{A}_L} \\ {\mathcal{\tilde{P}}_R \neq \mathcal{\tilde{P}}_S}}} \left(\prod_{i \in \eta_{s_1}
(\mathcal{P}_S,\mathcal{\tilde{P}}_S)}\left[\dfrac {1}{1+\dfrac{\left\vert \sigma_{sd}\left(x_{s_1}^i\left(\mathcal{P}_S\right)-x_{s_1}^i(\mathcal{\tilde{P}}_S)\right)\right\vert^2}{4}}\right] \hspace{200 cm}\right)\\
 \nonumber
& \hspace{5 cm} \prod_{i \in \left\lbrace\eta_{s_2}
(\mathcal{P}_S,\mathcal{\tilde{P}}_S) \cup \eta_{r}
(\mathcal{P}_R,\mathcal{\tilde{P}}_R)\right\rbrace}\left[\dfrac{1}{1+\dfrac{\left\vert \sigma_{sd}\left(x_{s_2}^i\left(\mathcal{P}_S\right)-x_{s_2}^i(\mathcal{\tilde{P}}_S)\right)\right\vert^2+\left\vert\sigma_{rd}\left(x_{r}^i\left(\mathcal{P}_R\right)-{x}_{r}^{i}(\mathcal{\tilde{P}}_R)\right)\right\vert^2}{4}}\right]\\
\label{eqn11_mod}
& \hspace{7 cm} \left. \prod_{i \in \left\lbrace\eta_{s_1}
(\mathcal{P}_S,\mathcal{{P}}_R) \cup \eta_{s_1}
(\mathcal{\tilde{P}}_S,\mathcal{\tilde{P}}_R)\right\rbrace}\left[\dfrac{1}{1+\dfrac{\left\vert \sigma_{sr}\left({x}_{s_1}^{i}(\mathcal{\tilde{P}}_R)-x_{s_1}^i(\mathcal{\tilde{P}}_S)\right)\right\vert^2 +\left\vert \sigma_{sr}\left({x}_{s_1}^i(\mathcal{P}_S)-x_{s_1}^i(\mathcal{{P}}_R)\right)\right\vert^2}{8}}\right]\right)
\end{align}
\hrule
\end{figure*}

\begin{proof} 
The PEP that a transmitted path $\mathcal{P}_S$ is decoded as ${\mathcal{\tilde{P}}_{S}}$ at D is

{\footnotesize
\begin{align}
\label{eqnmain}
Pr\left\lbrace\mathcal{P}_S\longrightarrow{\mathcal{\tilde{P}}_{S}}\right\rbrace = \sum_{\mathcal{P}_R\in \mathcal{A}_L}Pr\left\lbrace\mathcal{P}_S\longrightarrow{\mathcal{\tilde{P}}_{S}}\vert\mathcal{P}_R\right\rbrace Pr\left\lbrace\mathcal{P}_R\vert\mathcal{P}_S\right\rbrace,
\end{align}
}where $\mathcal{P}_R$ is the path to which R decodes.
The probability $Pr\left\lbrace{\mathcal{P}_R}\vert{\mathcal{P}_S}\right\rbrace$ is upper bounded by  \eqref{eqn2}.

%
 
 Also,
 {\footnotesize
 \begin{align}
 \nonumber
 Pr\left\lbrace\mathcal{P}_S\longrightarrow{\mathcal{\tilde{P}}_{S}}\vert\mathcal{P}_R\right\rbrace \\
 \label{eqn_con}
 &\hspace{-3 cm}= Pr\left\lbrace \min_{\mathcal{\tilde{P}}_R \in \mathcal{A}_L}\sum_{i=1}^{L}\phi_i\left(\mathcal{P}_S,\mathcal{\tilde{P}}_R\right) \geq \min_{\mathcal{\tilde{P}}_R \in \mathcal{A}_L}\sum_{i=1}^{L}\phi_i(\mathcal{\tilde{P}}_S,\mathcal{\tilde{P}}_R)\vert\left(\mathcal{P}_S,\mathcal{P}_R\right)\right\rbrace.
 \end{align}
 }
 
Let $\mathcal{F}_1$ and $\mathcal{F}_2$ be two sets defined as, 
 
 {\footnotesize
 \begin{align}
 \nonumber
 &\mathcal{F}_1=\left\lbrace \left(Y_{d_1}^{1,L},Y_{d_2}^{1,L}\right) \mid \min_{\mathcal{\tilde{P}}_R \in \mathcal{A}_L}\sum_{i=1}^{L}\phi_i\left(\mathcal{P}_S,\mathcal{\tilde{P}}_R\right)\right.\\
 \nonumber
 &\hspace{5 cm}\left.\geq \min_{\mathcal{\tilde{P}}_R \in \mathcal{A}_L}\sum_{i=1}^{L}\phi_i(\mathcal{\tilde{P}}_S,\mathcal{\tilde{P}}_R) \right\rbrace ,\\
\nonumber
&\mathcal{F}_2=\left\lbrace \left(Y_{d_1}^{1,L},Y_{d_2}^{1,L}\right) \mid \sum_{i=1}^L \phi_i\left(\mathcal{P}_S,\mathcal{P}_R\right)\geq \min_{\mathcal{\tilde{P}}_R \in \mathcal{A}_L}\sum_{i=1}^{L}\phi_i(\mathcal{\tilde{P}}_S,\mathcal{\tilde{P}}_R)\right\rbrace.
 \end{align}
 } 
 
 Since $\mathcal{F}_1 \subset \mathcal{F}_2$, from \eqref{eqn_con} we have,
 
 {\footnotesize
 \begin{align}
 \nonumber
 Pr\left\lbrace\mathcal{P}_S\longrightarrow{\mathcal{\tilde{P}}_{S}}\vert\mathcal{P}_R\right\rbrace \\
 \label{eqn_latest}
 &\hspace{-2 cm}\leq Pr\left\lbrace \sum_{i=1}^L \phi_i\left(\mathcal{P}_S,\mathcal{P}_R\right)\geq \min_{\mathcal{\tilde{P}}_R \in \mathcal{A}_L}\sum_{i=1}^{L}\phi_i(\mathcal{\tilde{P}}_S,\mathcal{\tilde{P}}_R)\vert\left(\mathcal{P}_S,\mathcal{P}_R\right)\right\rbrace
\end{align}
 }

 Let $\mathcal{F}_3$ be the set defined as,
 
 {\footnotesize
 \begin{align}
 \nonumber
 &\mathcal{F}_3=\bigcup_{\mathcal{\tilde{P}_R} \in \mathcal{A}_L}\left\lbrace \left(Y_{d_1}^{1,L},Y_{d_2}^{1,L}\right) \mid \sum_{i=1}^{L}\phi_i\left(\mathcal{P}_S,\mathcal{{P}}_R\right) \geq \sum_{i=1}^{L}\phi_i(\mathcal{\tilde{P}}_S,\mathcal{\tilde{P}}_R) \right\rbrace.
 \end{align}
 }
 
 From \eqref{eqn_latest}, noting that $\mathcal{F}_2 \subset \mathcal{F}_3$ and using the union bound, we have,
 
 {\footnotesize
 \begin{align}
\nonumber
Pr\left\lbrace\mathcal{P}_S\longrightarrow{\mathcal{\tilde{P}}_{S}}\vert\mathcal{P}_R\right\rbrace \\
 \label{eqn8}
 &\hspace{-2 cm}\leq \sum_{\mathcal{\tilde{P}_R}\in \mathcal{A}_L}Pr\left\lbrace \sum_{i=1}^L \phi_i\left(\mathcal{P}_S,\mathcal{P}_R\right)\geq \sum_{i=1}^{L} \phi_i\left(\mathcal{\tilde{P}}_S,\mathcal{\tilde{P}}_R\right) \vert \left(\mathcal{P}_S,\mathcal{P}_R\right)\right\rbrace.
 \end{align}
 }
 
The probability inside the summation in \eqref{eqn8} can be upper bounded using Lemma 1. The constant $c$, the vectors $x_i$, $i \in \lbrace 1,2 \rbrace$ and the vector $y$ , defined in Lemma 1, are chosen as given by \eqref{eqn_c_lemma1}-\eqref{y_lemma1} (shown at the top of the next page).

\begin{figure*}
{
\footnotesize
\begin{align}
\label{eqn_c_lemma1}
&c=\dfrac{1}{4} \sum_{i=1}^L\left(\left\vert h_{sr}^i\left({x}_{s_1}^{i}({\mathcal{{\tilde{P}}_R}})-x_{s_1}^i(\mathcal{\tilde{P}}_S)\right)\right\vert^2-\left\vert h_{sr}^i\left({x}_{s_1}^{i}({\mathcal{{{P}}_R}})-x_{s_1}^i(\mathcal{{P}}_S)\right)\right\vert^2\right)\\ 
\label{eqn_x1}
&x_1=[h_{sd_1}^1x_{s_1}^1(\mathcal{P}_S),h_{sd_1}^2x_{s_1}^2(\mathcal{P}_S),...,h_{sd_1}^L x_{s_1}^L(\mathcal{P}_S),
h_{sd_2}^1 x_{s_2}^1(\mathcal{P}_S)+h_{rd}^1 x_{rd}^1(\mathcal{P}_R),h_{sd_2}^2 x_{s_2}^2(\mathcal{P}_S)+h_{rd}^2 x_{rd}^2(\mathcal{P}_R),...,h_{sd_2}^L x_{s_2}^L (\mathcal{P}_S)+ h_{rd}^L x_{rd}^L(\mathcal{P}_R)]^T\\
\label{eqn_x2}
&x_2=[h_{sd_1}^1x_{s_1}^1(\mathcal{\tilde{P}}_S),h_{sd_1}^2x_{s_1}^2(\mathcal{\tilde{P}}_S),...,h_{sd_1}^L x_{s_1}^L(\mathcal{\tilde{P}}_S),
h_{sd_2}^1 x_{s_2}^1 (\mathcal{\tilde{P}}_S)+h_{rd}^1 x_{rd}^1 (\mathcal{\tilde{P}}_R),h_{sd_2}^2 x_{s_2}^2 (\mathcal{\tilde{P}}_S)+h_{rd}^2 x_{rd}^2 (\mathcal{\tilde{P}}_R),...,h_{sd_2}^L x_{s_2}^L (\mathcal{\tilde{P}}_S)+ h_{rd}^L x_{rd}^L(\mathcal{\tilde{P}}_R)]^T\\
\label{y_lemma1}
&y=[Y_{d_1}^1,Y_{d_1}^2,...,Y_{d_1}^L,Y_{d_2}^1,Y_{d_2}^2,...,Y_{d_2}^L]^T
\end{align}
\hrule
}
\end{figure*}
 
 Using Lemma \ref{lemma1} in \eqref{eqn8} gives rise to \eqref{eqn9} (\eqref{eqn9}-\eqref{eqn11} are shown on the next page). Substituting \eqref{eqn2} and \eqref{eqn9} in \eqref{eqnmain} gives \eqref{eqn10}. Taking expectation in \eqref{eqn10} with respect to $h_{sd_1}^i$, $h_{sd_2}^i$, $h_{rd}^i$ and $h_{sr}^i$, $1 \leq i \leq L$, yields \eqref{eqn11} which is the same as \eqref{eqn11_mod}. 
 
\begin{figure*}
 {\scriptsize
 \begin{align}
 \nonumber
 &Pr\left\lbrace\mathcal{P}_S\longrightarrow{\mathcal{\tilde{P}}_{S}}\vert\mathcal{P}_R\right\rbrace  \leq  \sum_{\mathcal{\tilde{P}}_R\in \mathcal{A}_L} \exp \left\lbrace \sum_{i=1}^L\left(-\dfrac{1}{2}\left\vert h_{sd_1}^i\left(x_{s_1}^i\left(\mathcal{P}_S\right)-x_{s_1}^i(\mathcal{\tilde{P}}_S)\right)\right\vert^2
 -\dfrac{1}{2}\left\vert h_{sd_2}^i\left(x_{s_2}^i\left(\mathcal{P}_S\right)-x_{s_2}^i(\mathcal{\tilde{P}}_S)\right)+h_{rd}^i\left(x_{r}^i\left(\mathcal{P}_R\right)-{x}_{r}^{i}(\mathcal{\tilde{P}}_R)\right)\right\vert^2\hspace{100 cm}\right)\right\rbrace\\
 \label{eqn9}
&\hspace{-25 cm}\left(\left\lbrace\hspace{33.5 cm}-\dfrac{1}{8}\left\vert h_{sr}^i\left({x}_{s_1}^{i}({\mathcal{{\tilde{P}}_R}})-x_{s_1}^i(\mathcal{\tilde{P}}_S)\right)\right\vert^2 +\dfrac{1}{8}\left\vert h_{sr}^i\left({x}_{s_1}^{i}({\mathcal{{{P}}_R}})-x_{s_1}^i(\mathcal{{P}}_S)\right)\right\vert^2
\right)\right\rbrace\\
\nonumber
 &Pr\left\lbrace\mathcal{P}_S\longrightarrow{\mathcal{\tilde{P}}_{S}}\right\rbrace  \leq  \sum_{\mathcal{{P}}_R\in \mathcal{A}_L}\sum_{\mathcal{\tilde{P}}_R\in \mathcal{A}_L} \exp \left\lbrace \sum_{i=1}^L\left(-\dfrac{1}{2}\left\vert h_{sd_1}^i\left(x_{s_1}^i\left(\mathcal{P}_S\right)-x_{s_1}^i(\mathcal{\tilde{P}}_S)\right)\right\vert^2
 -\dfrac{1}{2}\left\vert h_{sd_2}^i\left(x_{s_2}^i\left(\mathcal{P}_S\right)-x_{s_2}^i(\mathcal{\tilde{P}}_S)\right)+h_{rd}^i\left(x_{r}^i\left(\mathcal{P}_R\right)-{x}_{r}^{i}(\mathcal{\tilde{P}}_R)\right)\right\vert^2\hspace{100 cm}\right)\right\rbrace\\
 \label{eqn10}
&\hspace{-30 cm}\left(\left\lbrace\hspace{39.5 cm}-\dfrac{1}{8}\left\vert h_{sr}^i\left({x}_{s_1}^{i}(\mathcal{\tilde{P}}_R)-x_{s_1}^i(\mathcal{\tilde{P}}_S)\right)\right\vert^2 -\dfrac{1}{8}\left\vert h_{sr}^i\left({x}_{s_1}^i(\mathcal{P}_S)-x_{s_1}^i(\mathcal{{P}}_R)\right)\right\vert^2
\right)\right\rbrace\\
\nonumber
 &Pr\left\lbrace\mathcal{P}_S\longrightarrow{\mathcal{\tilde{P}}_{S}}\right\rbrace  \leq  \sum_{\mathcal{{P}}_R\in \mathcal{A}_L}\sum_{\mathcal{\tilde{P}}_R\in \mathcal{A}_L} \prod_{i=1}^{L}\left(\left[\dfrac {1}{1+\dfrac{\left\vert \sigma_{sd}\left(x_{s_1}^i\left(\mathcal{P}_S\right)-(x_{s_1}^i(\mathcal{\tilde{P}}_S)\right)\right\vert^2}{4}}\right]\left[\dfrac{1}{1+\dfrac{\left\vert \sigma_{sd}\left(x_{s_2}^i\left(\mathcal{P}_S\right)-(x_{s_2}^i(\mathcal{\tilde{P}}_S)\right)\right\vert^2+\left\vert\sigma_{rd}\left(x_{r}^i\left(\mathcal{P}_R\right)-{x}_{r}^{i}(\mathcal{\tilde{P}}_R)\right)\right\vert^2}{4}}\right] \hspace{200 cm}\right)\\
 \label{eqn11}
&\hspace{-20 cm}\left(\hspace{29 cm} \left[\dfrac{1}{1+\dfrac{\left\vert \sigma_{sr}\left({x}_{s_1}^{i}(\mathcal{\tilde{P}}_R)-x_{s_1}^i(\mathcal{\tilde{P}}_S)\right)\right\vert^2 +\left\vert \sigma_{sr}\left({x}_{s_1}^i(\mathcal{P}_S)-x_{s_1}^i(\mathcal{{P}}_R)\right)\right\vert^2}{8}}\right]\right)\\
\hline
\label{eqn12}
&Pr\left(\mathcal{P}_S \longrightarrow \mathcal{\tilde{P}_S}\right) \leq  {K} \prod_{i\in \eta_{s_1}(\mathcal{P}_S,\mathcal{\tilde{P}}_S)}\left[\dfrac {1}{{\left\vert \sigma_{sd}\left(x_{s_1}^i\left(\mathcal{P}_S\right)-(x_{s_1}^i(\mathcal{\tilde{P}}_S)\right)\right\vert^2}}\right]
\prod_{i\in \left\lbrace\eta_{s_2}(\mathcal{P}_S,\mathcal{\tilde{P}}_S) \cup \eta_{r}(\mathcal{P}_S,\mathcal{\tilde{P}}_S)\right\rbrace}\left[\dfrac{1}{{\left\vert \sigma_{sd}\left(x_{s_2}^i\left(\mathcal{P}_S\right)-(x_{s_2}^i(\mathcal{\tilde{P}}_S)\right)\right\vert^2+\left\vert\sigma_{rd}\left(x_{r}^i\left(\mathcal{P}_S\right)-{x}_{r}^{i}(\mathcal{\tilde{P}}_S)\right)\right\vert^2}}\right]
  \end{align}
  }
  \hrule
  \end{figure*} 

 \end{proof}
 \end{theorem}

The criteria to maximize the diversity order of the proposed TCM scheme is obtained in the following subsection. 
\subsection{Diversity Criteria}
\begin{figure}[htbp]
\centering
\includegraphics[totalheight=2in,width=3in]{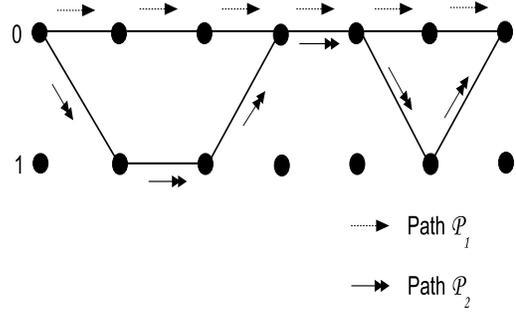}
\vspace{-0.5cm}	
\caption{Example illustrating the notion of unmerged length between two paths}	
\label{fig:path_example}
\end{figure}
\begin{definition}

The \textit{unmerged length} between the paths $\mathcal{P}_1$ and $\mathcal{{P}}_2$ in the trellis, denoted as $h(\mathcal{P}_1,\mathcal{{P}}_2)$ is the number of branches in which the paths $\mathcal{P}_1$ and $\mathcal{{P}}_2$ differ. For example, for the paths $\mathcal{P}_1$ and $\mathcal{P}_2$ of length 6 shown in Fig. \ref{fig:path_example}, the unmerged length $h(\mathcal{P}_1,\mathcal{{P}}_2)=5$. The \textit{unmerged length} of the code is the minimum value of $h(\mathcal{P}_1,\mathcal{{P}}_2)$, over all possible pairs of paths $(\mathcal{P}_1,\mathcal{{P}}_2)$ in the trellis.
\end{definition}
\begin{note}
Since the unmerged length between a pair of paths $\mathcal{P}_1,\mathcal{{P}}_2$ is the same in all three trellises $\mathcal{T}_{S_1}$, $\mathcal{T}_{S_2}$ and $\mathcal{T}_{R}$, we simply refer to it as the unmerged length in the trellis, without mentioning whether the trellis is $\mathcal{T}_{S_1}$, $\mathcal{T}_{S_2}$ or $\mathcal{T}_{R}$.
\end{note}

For two paths $\mathcal{P}_S$ and $\mathcal{\tilde{P}}_S$ whose unmerged length is equal to the unmerged length of the code, the first term in the PEP upper bound given by \eqref{eqn11_mod} is of diversity order upper bounded by twice the unmerged length of the code, since both $\vert\eta_{s_1}(\mathcal{P}_S,\mathcal{\tilde{P}}_S) \vert$ and $\vert\eta_{s_1}(\mathcal{P}_S,\mathcal{\tilde{P}}_S) \cup \eta_{r}(\mathcal{P}_S,\mathcal{\tilde{P}}_S) \vert$ are upper bounded by the unmerged length of the code.  Hence, the diversity order of the proposed TCM scheme is upper bounded by twice the unmerged length of the code. The mappings $\mathcal{X}_{S_1}$, $\mathcal{X}_{S_2}$ and $\mathcal{X}_{R}$ should be chosen such that this bound is met with equality.

 For two arbitrary paths $\mathcal{P}_S$ and $\mathcal{\tilde{P}}_S$ in the trellis, let the diversity order corresponding to the error event  $\left(\mathcal{P}_S \longrightarrow \mathcal{\tilde{P}_S}\right)$ be denoted as $D(\mathcal{P}_S ,\mathcal{\tilde{P}_S})$. 
 
 \begin{corollary}
 For the proposed TCM scheme, $D(\mathcal{P}_S ,\mathcal{\tilde{P}_S})$ is bounded as,
 
 {\footnotesize
\begin{align}
\nonumber
&\vert\eta_{s_1}(\mathcal{P}_S,\mathcal{\tilde{P}}_S)\vert + \vert\eta_{s_2}(\mathcal{P}_S,\mathcal{\tilde{P}}_S)\vert \leq D(\mathcal{P}_S ,\mathcal{\tilde{P}_S}) \\
\nonumber
&\hspace{1cm}\leq \min\left\lbrace\vert\eta_{s_1}(\mathcal{P}_S,\mathcal{\tilde{P}}_S)\vert+\vert\eta_{s_2}(\mathcal{P}_s,\mathcal{\tilde{P}}_S)\cup \eta_{r}(\mathcal{P}_S,\mathcal{\tilde{P}}_S)\vert,\right.\\
\label{div_order}
&\left. \hspace{4 cm}2\vert\eta_{s_1}(\mathcal{P}_S,\mathcal{\tilde{P}}_S)\vert+\vert\eta_{s_2}(\mathcal{P}_s,\mathcal{\tilde{P}}_S)\vert\right\rbrace,
\end{align}
}with equality on both the sides if and only if $\eta_{r}(\mathcal{P}_S,\mathcal{\tilde{P}}_S)$ is a subset of $\eta_{s_2}(\mathcal{P}_S,\mathcal{\tilde{P}}_S)$.

 \begin{proof}
 From  \eqref{eqn11_mod}, $D(\mathcal{P}_S ,\mathcal{\tilde{P}_S})$ is given by,
 
 {\footnotesize
 \begin{align}
 \nonumber
 D(\mathcal{P}_S ,\mathcal{\tilde{P}_S})&=\vert\eta_{s_1}(\mathcal{P}_S,\mathcal{\tilde{P}}_S)\vert 
 + \min_{\mathcal{P}_R,\mathcal{\tilde{P}}_R}\left\lbrace\vert\eta_{s_2}(\mathcal{P}_s,\mathcal{\tilde{P}}_S)\cup \eta_{r}(\mathcal{P}_R,\mathcal{\tilde{P}}_R)\vert\right.\\
 \nonumber
&\left.\hspace{3cm} +\vert\eta_{s_1}(\mathcal{\tilde{P}}_R,\mathcal{\tilde{P}}_S)\cup \eta_{s_1}(\mathcal{P}_R,\mathcal{{P}}_S)\vert\right\rbrace,\\
\nonumber
 &\geq \vert\eta_{s_1}(\mathcal{P}_s,\mathcal{\tilde{P}}_S)\vert + \min_{\mathcal{P}_R,\mathcal{\tilde{P}}_R}\vert\eta_{s_2}(\mathcal{P}_s,\mathcal{\tilde{P}}_S)\cup \eta_{r}(\mathcal{P}_R,\mathcal{\tilde{P}}_R)\vert\\
 \label{ineq1}
 &\hspace{2 cm}+\min_{\mathcal{P}_R,\mathcal{\tilde{P}}_R}\vert\eta_{s_1}(\mathcal{\tilde{P}}_R,\mathcal{\tilde{P}}_S)\cup \eta_{s_1}(\mathcal{P}_R,\mathcal{{P}}_S)\vert.
 \end{align}
 }
 
 The quantity $\vert\eta_{s_2}(\mathcal{P}_s,\mathcal{\tilde{P}}_S)\cup \eta_{r}(\mathcal{P}_R,\mathcal{\tilde{P}}_R)\vert$ attains the minimum value, for the case when $ \mathcal{P}_R=\mathcal{\tilde{P}}_R$, in which case $\eta_{r}(\mathcal{P}_R,\mathcal{\tilde{P}}_R)$ is an empty set, or for the case when the set $\eta_{r}(\mathcal{P}_R,\mathcal{\tilde{P}}_R)$ is a subset of $\eta_{s_2}(\mathcal{P}_s,\mathcal{\tilde{P}}_S)$. In both the cases the minimum value attained is $\vert\eta_{s_2}(\mathcal{P}_s,\mathcal{\tilde{P}}_S)\vert$.
 
 The minimum value of $\vert\eta_{s_1}(\mathcal{\tilde{P}}_R,\mathcal{\tilde{P}}_S)\cup \eta_{s_1}(\mathcal{P}_R,\mathcal{{P}}_S)\vert$ equal to zero is attained when $ \mathcal{P}_R=\mathcal{{P}}_S$ and $ \mathcal{\tilde{P}}_R=\mathcal{\tilde{P}}_S$.
 
 Hence from  \eqref{ineq1} we get the lower bound in \eqref{div_order}.
 
 The lower bound in \eqref{div_order} is met with equality if both the minima in \eqref{ineq1} are attained for the same choice of the pair $(\mathcal{P}_R,\mathcal{\tilde{P}}_R)$, which occurs if and only if $ \mathcal{P}_R=\mathcal{{P}}_S$, $\mathcal{\tilde{P}}_R=\mathcal{\tilde{P}}_S$ and the the set $\eta_{r}(\mathcal{P}_R,\mathcal{\tilde{P}}_R)$ is a subset of $\eta_{s_2}(\mathcal{P}_S,\mathcal{\tilde{P}}_S)$. In other words, equality in \eqref{ineq1} occurs if and only if $\eta_{r}(\mathcal{P}_S,\mathcal{\tilde{P}}_S)$ is a subset of $\eta_{s_2}(\mathcal{P}_S,\mathcal{\tilde{P}}_S)$.
 
 The choice $ \mathcal{P}_R=\mathcal{{P}}_S$ and $ \mathcal{\tilde{P}}_R=\mathcal{\tilde{P}}_S$ results in, 

{\footnotesize
 \begin{align}
 \nonumber
 &\left.\vert\eta_{s_2}(\mathcal{P}_s,\mathcal{\tilde{P}}_S)\cup \eta_{r}(\mathcal{P}_R,\mathcal{\tilde{P}}_R)\vert\right.\\
  \label{Div_ub1}
&\left.\hspace{0cm} +\vert\eta_{s_1}(\mathcal{\tilde{P}}_R,\mathcal{\tilde{P}}_S)\cup \eta_{s_1}(\mathcal{P}_R,\mathcal{{P}}_S)\vert\right.=\vert\eta_{s_2}(\mathcal{P}_s,\mathcal{\tilde{P}}_S)\cup \eta_{r}(\mathcal{P}_S,\mathcal{\tilde{P}}_S)\vert.
\end{align}
}

 The choice $ \mathcal{P}_R=\mathcal{\tilde{P}}_R=\mathcal{\tilde{P}}_S$ results in, 

{\footnotesize
 \begin{align}
 \nonumber
 &\left.\vert\eta_{s_2}(\mathcal{P}_s,\mathcal{\tilde{P}}_S)\cup \eta_{r}(\mathcal{P}_R,\mathcal{\tilde{P}}_R)\vert\right.\\
 \label{Div_ub2}
&\left.\hspace{0cm} +\vert\eta_{s_1}(\mathcal{\tilde{P}}_R,\mathcal{\tilde{P}}_S)\cup \eta_{s_1}(\mathcal{P}_R,\mathcal{{P}}_S)\vert\right.=\vert\eta_{s_2}(\mathcal{P}_s,\mathcal{\tilde{P}}_S)\vert+ \vert\eta_{s_1}(\mathcal{P}_S,\mathcal{\tilde{P}}_S)\vert.
\end{align}
}

From \eqref{Div_ub1} and \eqref{Div_ub2} we get the upper bound in \eqref{div_order}.

%

Equality in both the lower and upper bounds of \eqref{div_order} occurs if and only if $\eta_{r}(\mathcal{P}_S,\mathcal{\tilde{P}}_S)$ is a subset of $\eta_{s_2}(\mathcal{P}_S,\mathcal{\tilde{P}}_S)$.
\end{proof}
\end{corollary}

\begin{definition}
The \textit{effective length} of a pair of paths  $(\mathcal{P}_S,\mathcal{\tilde{P}}_S)$ in the trellis $\mathcal{T}_{S_1}$ is defined to be the cardinality of the set $\eta_{s_1}(\mathcal{P}_S,\mathcal{\tilde{P}}_S)$. The \textit{effective length} of the trellis $\mathcal{T}_{S_1}$ is defined to be minimum among the effective lengths of all possible pairs of paths  $(\mathcal{P}_S,\mathcal{\tilde{P}}_S)$. In a similar way, the effective lengths of the trellises $\mathcal{T}_{S_2}$ and $\mathcal{T}_{R}$ can be defined.
\end{definition}

\begin{definition}
The \textit{generalized effective length} of a pair of paths $(\mathcal{P}_S,\mathcal{\tilde{P}}_S)$ in the trellis pair $(\mathcal{T}_{S_2},\mathcal{T}_{R})$ is defined to be the cardinality of the set $\left\lbrace\eta_{s_2}(\mathcal{P}_S,\mathcal{\tilde{P}}_S) \cup \eta_{r}(\mathcal{P}_S,\mathcal{\tilde{P}}_S)\right\rbrace$.
\end{definition}

\begin{note}
If the labellings $\mathcal{X}_{s_2}$ and $\mathcal{X}_{r}$ are the same, the generalized effective length of the pair of paths reduces to the effective length of the pair of paths in the trellis $\mathcal{T}_{S_2}$, which is equal to the effective length of the pair of paths in the trellis $\mathcal{T}_{R}$.
\end{note}

If $\mathcal{X}_{s_1}$ and $\mathcal{X}_{s_2}$ are chosen such that the effective lengths of the trellises $\mathcal{T}_{S_1}$ and $\mathcal{T}_{S_2}$ are equal to the unmerged length of the code, then for every pair of paths $(\mathcal{P}_{S},\mathcal{\tilde{P}}_{S})$, $\vert \eta_{s_1}(\mathcal{P}_{S},\mathcal{\tilde{P}}_{S})\vert$ and $\vert \eta_{s_2}(\mathcal{P}_{S},\mathcal{\tilde{P}}_{S})\vert$ are greater than the unmerged length of the code. In that case, from the lower bound in Corollary 1 and from the fact that the diversity order cannot exceed twice the unmerged length of the code, it follows that the diversity order for the proposed scheme is equal to twice the unmerged length of the code. Hence, \emph{a sufficient condition to obtain maximum diversity order is to maximize the effective lengths of the trellises $\mathcal{T}_{S_1}$ and $\mathcal{T}_{S_2}$}, independent of the labelling $\mathcal{X}_r$. This is not surprising, since during Phase 2 the source encodes the same bits it encoded and transmitted during Phase 1, maximum diversity order is obtained irrespective of whether the relay transmits during Phase 2 or not. But since $\sigma_{sd}^2 \ll \sigma_{rd}^2$, from \eqref{eqn12} it can be seen that the coding gain is greatly reduced if the relay does not transmit during Phase 2. It is shown in the following subsection that the effective length of $\mathcal{T}_R$ also should equal the unmerged length of the code to avoid this reduction in  coding gain.

 Even though the above condition guarantees maximum diversity, there is no simple expression for $D(\mathcal{P}_S,\mathcal{\tilde{P}}_S)$ (Corollary 1 gives only the upper and lower bounds) and hence it is not clear which error events contribute to the minimum diversity order. Proposition 1 below gives the conditions under which an error event results in the minimum diversity order.

\begin{proposition}
Assuming that the sufficient condition to obtain full diversity is satisfied, the error event corresponding to a pair of paths contributes to the minimum diversity order if and only if the effective length of the pair of paths in $\mathcal{T}_{S_1}$ and the generalized effective length of the pair of paths in $(\mathcal{T}_{S_2},\mathcal{T}_{R})$, are equal to the unmerged length of the code.  
\begin{proof}
Let $\mathcal{P}_{S}$ and $\mathcal{\bar{P}}_{S}$ be two paths in the trellis. The maps $\mathcal{X}_{S_1}$ and $\mathcal{X}_{S_2}$ are chosen such that the effective lengths of the trellises $\mathcal{T}_{S_1}$ and $\mathcal{T}_{S_2}$ are equal to the unmerged length of the code. 

From the lower bound in Corollary 1, $D(\mathcal{P}_{S} ,\mathcal{\bar{P}}_{S})$ is greater than or equal to twice the unmerged length of the code, since $\vert\eta_{s_1}(\mathcal{P}_S,\mathcal{\bar{P}}_S)\vert$ and $\vert\eta_{s_2}(\mathcal{P}_S,\mathcal{\bar{P}}_S)\vert$ are greater than or equal to the unmerged length of the code. Equality can occur only when the set $\eta_{r}(\mathcal{P}_S,\mathcal{\bar{P}}_S)$ is a subset of $\eta_{s_2}(\mathcal{P}_S,\mathcal{\bar{P}}_S)$, and the effective lengths of the pair $(\mathcal{P}_S,\mathcal{\bar{P}}_S)$ in both the trellises $\mathcal{T}_{S_1}$ and $\mathcal{T}_{S_2}$, i.e., $\vert \eta_{s_1}(\mathcal{P}_S,\mathcal{\bar{P}}_S)\vert$ and $\vert \eta_{s_2}(\mathcal{P}_S,\mathcal{\bar{P}}_S) \vert$,  are equal to the unmerged length of the code. Clearly, $\eta_{r}(\mathcal{P}_S,\mathcal{\bar{P}}_S)$ is not a subset of $\eta_{s_2}(\mathcal{P}_S,\mathcal{\bar{P}}_S)$, if and only if $\vert \eta_{r}(\mathcal{P}_S,\mathcal{\bar{P}}_S) \cup \eta_{s_2}(\mathcal{P}_S,\mathcal{\bar{P}}_S) \vert \gneq \vert\eta_{s_2}(\mathcal{P}_S,\mathcal{\bar{P}}_S)\vert$, in which case the generalized effective length of the pair of paths $(\mathcal{P}_S,\mathcal{\bar{P}}_S)$ in the trellis pair $(\mathcal{T}_{S_2},\mathcal{T}_{R})$ is greater than unmerged length of the code. Hence, the error event corresponding to a pair of paths give rise to the minimum diversity order, if and only if the conditions given in the statement of the Proposition are satisfied.  
\end{proof}
\end{proposition}

In the rest of the paper, it is assumed that the sufficient condition to obtain maximum diversity is satisfied.

For  a pair of paths $\mathcal{P}_{S}$ and $\mathcal{\tilde{P}}_{S}$ resulting in the minimum diversity, only the first term in \eqref{eqn11_mod} gives rise to minimum diversity order. Neglecting terms of higher diversity order in \eqref{eqn11_mod}, we get \eqref{eqn12} (shown on the previous to the previous page), where $K$ is a positive constant.

The criterion to maximize the coding gain is obtained in the following subsection.
\subsection{Coding Gain Criterion}

 
Let $m_1(\mathcal{P}_S,\mathcal{\tilde{P}}_S)$ denote the product distance of the pair of paths $(\mathcal{P}_S,\mathcal{\tilde{P}}_S)$ in the trellis $\mathcal{T}_{S_1}$, i.e., 

{\footnotesize
\begin{align}
\nonumber
&m_1(\mathcal{P}_S,\mathcal{\tilde{P}}_S)=\prod_{i\in \eta_{s_1}(\mathcal{P}_S,\mathcal{\tilde{P}}_S)}{\left\vert x_{s_1}^i\left(\mathcal{P}_S\right)-x_{s_1}^i(\mathcal{\tilde{P}}_S)\right\vert^2}.
\end{align}
}

\begin{definition}
The \textit{generalized product distance} of the pair of paths $(\mathcal{P}_S,\mathcal{\tilde{P}}_S)$ corresponding to the trellis pair $(\mathcal{T}_{S_2},\mathcal{T}_R)$ denoted as $m_2(\mathcal{P}_S,\mathcal{\tilde{P}}_S)$ is defined as,

{\footnotesize
\begin{align}
\nonumber
&m_2(\mathcal{P}_S,\mathcal{\tilde{P}}_S)=\prod_{i\in \left\lbrace\eta_{s_2}(\mathcal{P}_S,\mathcal{\tilde{P}}_S) \cup \eta_{r}(\mathcal{P}_S,\mathcal{\tilde{P}}_S)\right\rbrace}\left(\gamma\left\vert x_{s_2}^i\left(\mathcal{P}_S\right)-x_{s_2}^i(\mathcal{\tilde{P}}_S)\right\vert^2\hspace{100 cm}\right)\\
\label{gpd}
&\hspace{-20 cm}\left(\hspace{25 cm}+\left\vert x_{r}^i\left(\mathcal{P}_S\right)-{x}_{r}^{i}(\mathcal{\tilde{P}}_S)\right\vert^2\right),
\end{align} 
 }where $\gamma=\sigma_{sd}^2/\sigma_{rd}^2$.

\end{definition}

\begin{note}
If the labelling used for the trellises $\mathcal{T}_{R}$ and $\mathcal{T}_{S_2}$ are the same, the generalized product distance  of the pair of paths $(\mathcal{P}_S,\mathcal{\tilde{P}}_S)$ corresponding to the trellis pair $(\mathcal{T}_{S_2},\mathcal{T}_R)$, reduces within a constant scaling factor to the product distance of the pair of paths $(\mathcal{P}_S,\mathcal{\tilde{P}}_S)$ in the trellis $\mathcal{T}_{R}$, which is equal to the product distance of the pair of paths $(\mathcal{P}_S,\mathcal{\tilde{P}}_S)$ in the trellis $\mathcal{T}_{S_2}$.
\end{note}

\begin{definition}
 Consider the trellis triplet $(\mathcal{T}_{S_1},\mathcal{T}_{S_2},\mathcal{T}_R)$. The combined product distance of the pair of paths $(\mathcal{P}_S,\mathcal{\tilde{P}}_S)$, denoted by $m(\mathcal{P}_S,\mathcal{\tilde{P}}_S)$, is defined as the product of the product distance of the pair of paths $(\mathcal{P}_S,\mathcal{\tilde{P}}_S)$ in the trellis $\mathcal{T}_{S_1}$ and the generalised product distance of the pair of paths $(\mathcal{P}_S,\mathcal{\tilde{P}}_S)$ corresponding to the trellis pair $(\mathcal{T}_{S_2},\mathcal{T}_R)$, i.e., 

{\footnotesize
\begin{align}
 \nonumber
 &{m(\mathcal{P}_S,\mathcal{\tilde{P}}_S)}= \left[m_1(\mathcal{P}_S,\mathcal{\tilde{P}}_S)m_{2}(\mathcal{P}_S,\mathcal{\tilde{P}}_S)\right].
\end{align}
}

\end{definition}

\begin{definition}
The combined product distance of the trellis triplet $(\mathcal{T}_{S_1},\mathcal{T}_{S_2},\mathcal{T}_R)$, which is also the coding gain metric denoted by $\mathcal{G}$, is defined to be the minimum value of $m(\mathcal{P}_S,\mathcal{\tilde{P}}_S)$, 
i.e.,
%
{\footnotesize
\begin{align}
 \nonumber
 &\mathcal{G}= \min_{(\mathcal{P}_S,\mathcal{\tilde{P}}_S) \in \mathcal{Z_P}}\left[m(\mathcal{P}_S,\mathcal{\tilde{P}}_S)\right].
\end{align}
}where $\mathcal{Z_P}$ denotes the set of all pairs of paths $(\mathcal{P}_S,\mathcal{\tilde{P}}_S)$ whose effective length in $\mathcal{T}_{S_1}$ and the generalized effective length in $(\mathcal{T}_{S_2},\mathcal{T}_R)$ are equal to the unmerged length of the code.
\end{definition}

From \eqref{eqn12}, it can be seen that \emph{the combined product distance of the trellis triplet $(\mathcal{T}_{S_1},\mathcal{T}_{S_2},\mathcal{T}_R)$ needs to be maximized, to maximize the coding gain}. For a pair of paths $(\mathcal{P}_S,\mathcal{\tilde{P}}_S)$ which contribute to the minimum diversity order, if the effective length of $(\mathcal{P}_S,\mathcal{\tilde{P}}_S)$ in the trellis $\mathcal{T}_R$ is not equal to the unmerged length of the code, then $x_r^i(\mathcal{P}_S)=x_r^i(\mathcal{\tilde{P}}_S)$ for some $i \in \eta_r(\mathcal{P}_S,\mathcal{\tilde{P}}_S)$. In that case $m_2(\mathcal{P}_S,\mathcal{\tilde{P}}_S)$ is greatly reduced since $\gamma \ll 1$. In order to avoid this, the effective length of the trellis $\mathcal{T}_R$ should be made equal to the unmerged length of the code.

\subsection{Near Optimality of the Proposed Near-ML decoder}

\begin{figure*}
\scriptsize
\begin{align}
\label{ideal_PEP}
 &Pr\left\lbrace\mathcal{P}_S\longrightarrow{\mathcal{\tilde{P}}_{S}}\right\rbrace  \leq   \exp \left\lbrace \sum_{i=1}^L\left(-\dfrac{1}{2}\left\vert h_{sd_1}^i\left(x_{s_1}^i\left(\mathcal{P}_S\right)-(x_{s_1}^i(\mathcal{\tilde{P}}_S)\right)\right\vert^2
 -\dfrac{1}{2}\left\vert h_{sd_2}^i\left(x_{s_2}^i\left(\mathcal{P}_S\right)-(x_{s_2}^i(\mathcal{\tilde{P}}_S)\right)+h_{rd}^i\left(x_{r}^i\left(\mathcal{P}_S\right)-{x}_{r}^{i}(\mathcal{\tilde{P}}_S)\right)\right\vert^2\right)\right\rbrace\\
 \nonumber
 &Pr\left(\mathcal{P}_S \longrightarrow \mathcal{\tilde{P}_S}\right) \leq  {K} \prod_{i\in \eta_{s_1}(\mathcal{P}_S,\mathcal{\tilde{P}}_S)}\left[\dfrac {1}{{\left\vert \sigma_{sd}\left(x_{s_1}^i\left(\mathcal{P}_S\right)-(x_{s_1}^i(\mathcal{\tilde{P}}_S)\right)\right\vert^2}}\right]\\
 \label{ideal_PEP_avg}
&\hspace{5 cm}\prod_{i\in\left\lbrace\eta_{s_2}(\mathcal{P}_S,\mathcal{\tilde{P}}_S) \cup \eta_{r}(\mathcal{P}_S,\mathcal{\tilde{P}}_S)\right\rbrace}\left[\dfrac{1}{{\left\vert \sigma_{sd}\left(x_{s_2}^i\left(\mathcal{P}_S\right)-(x_{s_2}^i(\mathcal{\tilde{P}}_S)\right)\right\vert^2+\left\vert\sigma_{rd}\left(x_{r}^i\left(\mathcal{P}_S\right)-{x}_{r}^{i}(\mathcal{\tilde{P}}_S)\right)\right\vert^2}}\right]
  \end{align}
  \hrule
  \end{figure*} 
  
The following argument proves the high SNR near optimality of the proposed near-ML decoder for the TCM scheme. Consider the situation where the S-R link is ideal, i.e., the relay decodes all the bits correctly. The optimal ML decoder at D decides in favour of the path given by,

{\footnotesize 
 \begin{align}
 \nonumber
 \mathcal{\hat{P}}_S&=\arg \min _{\mathcal{P}_S\in \mathcal{A}_L} \left(\sum_{i=1}^{L}\left\vert Y_{d_1}^i-h_{sd_1}^i x_{s_1}^i\left(\mathcal{P}_S\right)\right\vert^2 \right.\\
 \nonumber
 &\hspace{2.5 cm}\left.+ \left\vert Y_{d_2}^i-h_{sd_2}^i x_{s_2}^i\left(\mathcal{P}_S\right)-h_{rd}^i x_{r}^i\left(\mathcal{P}_S\right)\right\vert^2\right).
 \end{align}
 }
 
For this case, the PEP that the path $\mathcal{P}_S$ is decoded as $\mathcal{\tilde{P}}_S$ is upper bounded by \eqref{ideal_PEP} (shown at the top of the next page). Taking expectation with respect to the fading coefficients, we get \eqref{ideal_PEP_avg} (shown at the top of the next page). From \eqref{eqn12} and \eqref{ideal_PEP_avg}, we see that the high SNR bounds on the PEP for the proposed near ML decoder with a non-ideal S-R link is same as that of the optimal ML decoder with an ideal S-R link, if we consider only the error events giving rise to the minimum diversity order. The simulation results presented in Section VII confirm that the BER vs SNR performance of the proposed near ML decoder with a non-ideal S-R link approaches the performance of optimal ML decoder with an ideal S-R link, at high SNR. The high SNR performance of the optimal ML decoder for the proposed TCM scheme, with a non-ideal S-R link cannot be better than that of optimal ML decoder for the case when the S-R link is ideal. This implies the high SNR performance of the proposed near-ML decoder approaches the performance of the optimal ML decoder.

 From \eqref{eqn12}, it is clearly seen that even for the uncoded transmission scheme, choosing the labelling schemes used at S and R properly can potentially yield a significant performance improvement. The effect of the choice of labelling on the performance for the uncoded transmission scheme is discussed in the following section.  
\section{THE UNCODED TRANSMISSION SCHEME}
Consider the uncoded transmission scheme in which bits are directly mapped onto complex symbols at S and R. 
A collection of $log_2M$ bits constitutes a message. Let $\mathcal{M}=\lbrace1,2,...,M\rbrace$ denote this message set. 
The uncoded transmission scheme has an equivalent one state trellis representation, with $M$ edges connecting two successive stages. The unmerged length between every pair of distinct paths is one and  hence no distinction needs to be made between paths and edges. The index $i$ in the transmitted complex numbers $x_{s_1}^i(.)$, $x_{s_2}^i(.)$ and $x_{r}^i(.)$ can be dropped and they are the same as the labellings on the edges $\mathcal{X}_{s_1}(.)$, $\mathcal{X}_{s_2}(.)$ and $\mathcal{X}_{r}(.)$. Each path (edge) in the trellis is identified by the message $a \in \mathcal{M}$ which gets transmitted. 

\begin{corollary}
For the uncoded transmission scheme, the PEP that the decoder at D decides in favour of message $\bar{a} \in \mathcal{M}$ given that the message transmitted by the source was $a \in \mathcal{M}$ is upper bounded as,

{\footnotesize
\begin{align}
\nonumber
&Pr \left(a \longrightarrow \bar{a} \right) \leq \\
\nonumber
&\hspace{0 cm} \left [\dfrac{1}{1+\dfrac{1}{4}\vert \sigma_{sd}\vert ^2\vert x_{s_1}({a})-x_{s_1}(\bar{a})\vert ^2}\right]\\
\nonumber
&\left[\dfrac{1}{1+\dfrac{1}{4}\vert \sigma_{sd}\vert ^2\vert x_{s_2}({a})-x_{s_2}(\bar{a})\vert ^2+\dfrac{1}{4}\vert \sigma_{rd}\vert ^2\vert x_{r}({a})-x_{r}(\bar{a})\vert ^2}\right]\\
\nonumber
&\hspace{7.7 cm} + H.O.T.
\end{align}
}where H.O.T denotes the terms of diversity order greater than 2.
 
\begin{proof}
From \eqref{eqn11_mod}, replacing the paths $\mathcal{P}_S$, $\mathcal{\tilde{P}}_S$, $\mathcal{P}_R$ and $\mathcal{\tilde{P}}_R$ by the corresponding message indices, we get \eqref{eqn_PEP_uncoded} (shown at the top of the next page). Neglecting terms of diversity order greater than 2 gives the result.
\end{proof} 
\end{corollary}

\begin{figure*}
\scriptsize
\begin{align}
\nonumber
&Pr(a \longrightarrow \bar{a}) \leq  \sum_{l=1}^M\left(\left[\dfrac{1}{1+\dfrac{\vert \sigma_{ds}
\left(x_{s_1}\left(a\right)-x_{s_1}\left(\bar{a}\right)\right)\vert^2}{4}}\right]
\left[\dfrac{1}{1+\dfrac{\vert \sigma_{ds}
\left(x_{s_2}\left(a\right)-x_{s_2}\left(\bar{a}\right)\right)\vert^2}{4} + \dfrac{\vert \sigma_{dr}
\left(x_{r}\left(a\right)-x_{r}\left({l}\right)\right)\vert^2}{4}}\right]\hspace{30 cm}\right)\\
\nonumber
&\hspace{-15 cm}\left(\hspace{27.8 cm}\left[\dfrac{1}{1+\dfrac{\vert \sigma_{rs}
\left(x_{s_1}\left(\bar{a}\right)-x_{s_1}\left(l\right)\right)\vert^2}{8}}\right]\hspace{0 cm}\right)\\
\nonumber
&\hspace{2 cm}+\sum_{j=1,j \neq a}^{M}\sum_{m=1}^M\left(\left[\dfrac{1}{1+\dfrac{\vert \sigma_{ds}
\left(x_{s_1}\left(a\right)-x_{s_1}\left(\bar{a}\right)\right)\vert^2}{4}}\right]
\left[\dfrac{1}{1+\dfrac{\vert \sigma_{ds}
\left(x_{s_2}\left(a\right)-x_{s_2}\left(\bar{a}\right)\right)\vert^2}{4} + \dfrac{\vert \sigma_{dr}
\left(x_{r}\left(j\right)-x_{r}\left({m}\right)\right)\vert^2}{4}}\right]\hspace{30 cm}\right)\\
\label{eqn_PEP_uncoded}
&\hspace{-20 cm}\left(\hspace{29 cm}\left[\dfrac{1}{1+\dfrac{\vert \sigma_{rs}
\left(x_{s_1}\left({a}\right)-x_{s_1}\left(j\right)\right)\vert^2}{8}+\dfrac{\vert \sigma_{rs}
\left(x_{s_1}\left(\bar{a}\right)-x_{s_1}\left(m\right)\right)\vert^2}{8}}\right]\right)
\end{align}
\hrule
\end{figure*}

From Corollary 2, it follows that the diversity order of the uncoded transmission scheme is two and in order to minimise the PEP, we need to maximize the following:

{\footnotesize
\begin{align}
\nonumber
&\hspace{0 cm} \left [{1+\dfrac{1}{4}\vert \sigma_{sd}\vert ^2\vert x_{s_1}({a})-x_{s_1}(\bar{a})\vert ^2}\right]\\
\nonumber
&\left[{1+\dfrac{1}{4}\vert \sigma_{sd}\vert ^2\vert x_{s_2}({a})-x_{s_2}(\bar{a})\vert ^2+\dfrac{1}{4}\vert \sigma_{rd}\vert ^2\vert x_{r}({a})-x_{r}(\bar{a})\vert ^2}\right].
\end{align}
}

At high SNR we need to maximize the metric,

{\footnotesize
\begin{align}
\nonumber
m(a,\bar{a})=&{\vert x_{s_1}({a})-x_{s_1}(\bar{a})\vert ^2}\\
\nonumber
&\left[{\gamma\vert x_{s_2}({a})-x_{s_2}(\bar{a})\vert ^2+\vert x_{r}({a})-x_{r}(\bar{a})\vert ^2}\right],
\end{align}
}over all message pairs $(a,\bar{a})$.



By a labelling scheme, we refer to the triplet $(\mathcal{X}_{s_1},\mathcal{X}_{s_2},\mathcal{X}_{r})$. Let $\mathcal{L}$ denote a labelling scheme used at S and R. 
Let us define,
\begin{align}
\nonumber
d(\mathcal{L})&=\min_{a,\bar{a}, a \neq \bar{a}} m(a,\bar{a}),
\end{align}where $a,\bar{a} \in \mathcal{M}$.

Let $\mathcal{L}_0$ denote the labelling scheme in which the mapping from bits to complex symbols, used by S (during Phase 1 and Phase 2) and R (during Phase 2), are the same. Similar to $d(\mathcal{L})$, $d(\mathcal{L}_0)$ can be defined for the labelling scheme $\mathcal{L}_0$.
\begin{definition}
The Labelling Gain of the labelling scheme $\mathcal{L}$, which is a measure of the performance gain provided by $\mathcal{L}$ over $\mathcal{L}_0$, is given by,
\newline
\center
$L_G(\mathcal{L})=10\log_{10}\left[{\dfrac{d(\mathcal{L})}{d(\mathcal{L}_0)}}\right]$ dB.
\end{definition}
It is important to note that the Labelling Gain is calculated based on the upper bound on the PEP, taking into consideration only those pair of messages $a$ and $\bar{a}$ which contribute dominantly to the metric $m(a,\bar{a})$. The actual high SNR gain provided by the labelling scheme  $\mathcal{L}$ over the scheme $\mathcal{L}_0$ need not equal $L_G(\mathcal{L})$.

Throughout, the phrase \textit{with our labelling} means that S and R use the labelling scheme which is described in the following subsection and  \textit{with constant labelling} means that S and R use the labelling scheme $\mathcal{L}_0$.
\subsection{A Labelling Scheme for PSK Constellation}

In this subsection, we provide a good labelling scheme for $2^l$-PSK constellation, where $l \geq 2$ is an integer.
Let $s_{k}=\exp{\left(j k 2 \pi / M\right)} $, where $0 \leq k \leq M-1$ and $ M = 2^l$, denote the signal points in the $2^l$-PSK signal set.

Let $\mathcal{\bar{L}}$ denote the labelling scheme described as follows:
The maps $\mathcal{X}_{s_1}$, $\mathcal{X}_{r}$ and $\mathcal{X}_{s_2}$ are given by,

{\footnotesize
\begin{align}
\nonumber
&\mathcal{X}_{s_1}(k)=s_k,\\
\nonumber
&\mathcal{X}_{r}(k)=
\left\lbrace
\begin{array}{ll}
\nonumber
s_k, &\: \mathrm{if}  \: k \: \mathrm{is} \: \mathrm{even},\\
\nonumber
s_{(k+M/2 \mod M)}, &\: \mathrm{if} \: k \: \mathrm{is} \: \mathrm{odd},\\
\end{array}
\right.\\
\nonumber
&\mathcal{X}_{s_2}(k)=s_k,
\end{align}
}where $0 \leq k \leq  M-1$. 

\begin{theorem}
For the labelling scheme $\mathcal{\bar{L}}$, with $2^l$-PSK signal set, the labelling gain, assuming $\gamma \ll 1$, is approximately given by,


{\footnotesize
\begin{align}
\nonumber
{L}_G(\mathcal{\bar{L}})
\nonumber
&\approx \left\lbrace
\begin{array}{ll}
\nonumber
&20\log_{10}\left[\cot(\pi/2^l)\right], \: \mathrm{for}  \: l=2,3;\\
\nonumber
&20\log_{10}\left[4 \cos^2(\pi/2^l)\right], \: \mathrm{for}  \: l \geq 4.\\
\end{array}
\right.\\
\label{eqn_final}
\end{align}
}

\begin{proof}
Let us define,
{\footnotesize
\begin{align}
\nonumber
d_1(k,k')&=\vert \mathcal{X}_{s_1}(k)-\mathcal{X}_{s_1}(k')\vert \\
\nonumber
&= \vert \exp(jk2\pi/M)-\exp(jk'2 \pi /M ) \vert \\
\nonumber
&= \vert 1-\exp(j(k'-k)2 \pi /M )\vert ,
\end{align}
}where $0 \leq k , k' \leq M-1$.

Since $d_1(k,k')$ depends only on $\vert k-k'\vert$, let us denote it by $d_1(n)$, where $0 \leq \vert k-k' \vert =n \leq M$ . Then,

{\footnotesize
\begin{align}
\nonumber
d_1(n)&= \vert 1-\exp(jn2 \pi /M )\vert \\
\label{distance1}
&= 2 \vert\sin(\pi n / M)\vert,
 \end{align}
}where $0 \leq n \leq M$. Similarly, let us define,

{\footnotesize
\begin{align}
\nonumber
d_2(n)&=\vert \mathcal{X}_{r}(k)-\mathcal{X}_{r}(k')\vert\\
\nonumber
&=\left\lbrace
\begin{array}{ll}
&\vert 1-\exp(jn2 \pi /M ) \vert , \: \mathrm{for}  \: n \: \mathrm{even}\\
\nonumber
&\vert 1+\exp(jn2 \pi /M ) \vert , \: \mathrm{for}  \: n \: \mathrm{odd}\\
\end{array}
\right.\\
&=\left\lbrace
\begin{array}{ll}
&2\vert \sin( \pi n /M ) \vert , \: \mathrm{for}  \: n \: \mathrm{even}\\
\label{distance2}
&2\vert\cos( \pi n /M ) \vert , \: \mathrm{for}  \: n \: \mathrm{odd}\\
\end{array}
\right..
\end{align}
}

Since $\gamma \ll 1$, we have

{\footnotesize
\begin{align}
\nonumber
d(\mathcal{\bar{L}}) &\approx \min_{k \neq k'}\vert \mathcal{X}_{s_1}({k})-\mathcal{X}_{s_1}(k')\vert ^2\vert \mathcal{X}_{r}({k})-\mathcal{X}_{r}({k'})\vert ^2\\
\nonumber
&=\min_{1 \leq n \leq M} d_1^2(n)d_2^2(n)\\
\label{distance}
&=\min_{1 \leq n \leq M}(\min_{n \: \textrm{even}} [d_1^2(n)d_2^2(n)], \min_{n \: \textrm{odd}} [d_1^2(n)d_2^2(n)]).
\end{align}
}

From \eqref{distance1} and \eqref{distance2} we have, 

{\footnotesize
\begin{align}
\nonumber
\min_{\substack{ {1 \leq n \leq M} \\{n \: \textrm{even}}}} [d_1^2(n)d_2^2(n)] &= \min_{\substack{ {1 \leq n \leq M} \\{n \: \textrm{even}}}} [16 \sin^4(\pi n/M)]\\
\label{distance3}
&=16\sin^4(2 \pi /M)
\end{align}}
and similarly,

{\footnotesize
\begin{align}
\nonumber
\min_{\substack{ {1 \leq n \leq M} \\{n \: \textrm{odd}}}} [d_1^2(n)d_2^2(n)] &= \min_{\substack{ {1 \leq n \leq M} \\{n \: \textrm{odd}}}} [16 \sin^2(\pi n/M) \cos^2(\pi n/M)]\\
\label{distance4}
&=4\sin^2(2 \pi /M).
\end{align}}

Substituting \eqref{distance3} and \eqref{distance4} in \eqref{distance} we have,

{\footnotesize
\begin{align}
\label{distance5}
d(\mathcal{\bar{L}}) &\approx \min[16 \sin^4(2 \pi /M), 4 \sin^2(2 \pi /M)].
\end{align}}

Also, we have

{\footnotesize
\begin{align}
\nonumber
d(\mathcal{{L}}_0) &\approx \min_{1 \leq n \leq  M} d_1^4(n)\\
\label{distance6}
&=16 \sin^4(\pi /M).
\end{align}
}

Hence, from \eqref{distance5} and \eqref{distance6}, by definition of ${L}_G(\mathcal{\bar{L}})$, we have,

{\footnotesize
\begin{align}
\nonumber
{L}_G(\mathcal{\bar{L}})=20\log_{10}[\min(\cot(\pi/M),4 \cos^2(\pi/M))].
\end{align}
}

It can be verified that $\cot(\pi/M) \gneq 4 \cos^2(\pi/M)$ for $M \gneq 12$, from which we get \eqref{eqn_final}.
\end{proof}

\end{theorem}

 It can be seen from Theorem 2 that for $l=2$, the labelling gain ${L}_G(\mathcal{\bar{L}}) \approx 0$ dB. In other words, for the uncoded transmission scheme using 4 PSK, using different labelling schemes at S and R does not provide any performance improvement. It can be seen that $\cot(\pi/2^l)$ for $l = 3$,  and  $4\cos^2(\pi/2^l)$ for $l \geq 4$ are greater than one, whence the approximate value of ${L}_G(\mathcal{\bar{L}})$ is greater than 0 dB for $l \geq 3$. Since ${L}_G(\mathcal{\bar{L}})$ is always greater than the approximate value given by Theorem 2, for the uncoded transmission scheme using $2^l$ PSK, using our labelling scheme provides advantage over constant labelling, for all $l \geq 3$. Furthermore, since $4\cos^2(\pi/2^l)$ is an increasing function of $l$, the labelling gain increases with increasing size of the PSK constellation.
Simulation results presented in Section VII confirm that the uncoded transmission with our labelling outperforms the uncoded transmission with constant labelling.
 \section{TCM CODE DESIGN GUIDELINES AND EXAMPLES}
 
 \subsection{TCM Code Design Guidelines}

Let the generalized product distance $m_{2}(\mathcal{P}_S,\mathcal{\tilde{P}}_S)$  be split into two parts as,

 {\footnotesize 
 \begin{align}
 \nonumber
 m_2(\mathcal{P}_S,\mathcal{\tilde{P}}_S)=m_{21}(\mathcal{P}_S,\mathcal{\tilde{P}}_S)+m_{22}(\mathcal{P}_S,\mathcal{\tilde{P}}_S),
\end{align} 
 }where the product distances $m_{21}(\mathcal{P}_S,\mathcal{\tilde{P}}_S)$ and $m_{22}(\mathcal{P}_S,\mathcal{\tilde{P}}_S)$ are given by,
 
{\footnotesize
 \begin{align}
 \nonumber 
 &m_{21}(\mathcal{P}_S,\mathcal{\tilde{P}}_S)=\prod_{i\in \eta_2(\mathcal{P}_S,\mathcal{\tilde{P}}_S)}\left\vert x_{r}^i\left(\mathcal{P}_S\right)-{x}_{r}^{i}(\mathcal{\tilde{P}}_S)\right\vert^2,\\
 \nonumber
& m_{22}(\mathcal{P}_S,\mathcal{\tilde{P}}_S)=\prod_{i\in \eta_2(\mathcal{P}_S,\mathcal{\tilde{P}}_S)}\gamma\left\vert x_{s_2}^i\left(\mathcal{P}_S\right)-x_{s_2}^i(\mathcal{\tilde{P}}_S)\right\vert^2.
\end{align}
}

Let $\mathcal{G}_1$ and $\mathcal{G}_2$ be two metrics defined as,

{\footnotesize
\begin{align}
\nonumber
&\mathcal{G}_1= \min_{(\mathcal{P}_S,\mathcal{\tilde{P}}_S) \in \mathcal{Y_{P}}}\left[m_1(\mathcal{P}_S,\mathcal{\tilde{P}}_S)\right],\\
\nonumber
&\mathcal{G}_2= \min_{(\mathcal{P}_S,\mathcal{\tilde{P}}_S) \in \mathcal{Z_P}}\left[m_1(\mathcal{P}_S,\mathcal{\tilde{P}}_S)m_{21}(\mathcal{P}_S,\mathcal{\tilde{P}}_S)\right],
\end{align}
}where $\mathcal{Y_{P}}$ denotes the set of pairs of paths whose effective length is equal to the  effective length of the trellis $\mathcal{T}_{S_1}$.
 
  The code design guidelines are summarized below. Some of the design rules for TCM for the point to point AWGN and fading channels (\cite{Un},\cite{Da}), carry over for the relay channel as well.
 \begin{itemize}
 \item
  Since trellises with parallel transitions limit the effective length of the trellis to one, they should be avoided.
 \item
 Signal points from the signal set should occur with equal frequency.
 \item  
 In a trellis with regularity and symmetry, two branches which emerge from the same state form a part of a pair of paths which differ by a unmerged length equal to the unmerged length of the code. Since the coding gain metric $\mathcal{G}$ involves parameters $m_1(\mathcal{P}_S,\mathcal{\tilde{P}}_S)$, $m_{21}(\mathcal{P}_S,\mathcal{\tilde{P}}_S)$ and $m_{22}(\mathcal{P}_S,\mathcal{\tilde{P}}_S)$, which are of the form of product of the branch Euclidean distances of the pair of paths $(\mathcal{P}_S,\mathcal{\tilde{P}}_S)$, signal points assigned to branches emerging from the same state should be from the same Ungerboeck partition \cite{Un}.
\end{itemize}

Since $\gamma \ll 1$, the values of $\mathcal{G}$ and $\mathcal{G}_2$ are nearly equal and instead of maximizing $\mathcal{G}$, $\mathcal{G}_2$ can be maximized. In the design examples presented, the choice of $\mathcal{X}_{s_1}$ is made such that the effective length of $\mathcal{T}_{S_1}$ is maximized and the value of $\mathcal{G}_1$ is made large. The choice of $\mathcal{X}_{s_2}$ and $\mathcal{X}_{r}$ are made such that the effective length of $\mathcal{T}_{S_2}$ is maximized and the value of $\mathcal{G}_2$ is made large.  In the design examples presented, while the diversity order obtained is maximum, it is not claimed that the value of $\mathcal{G}_2$ for all the cases is maximum. As was the case with TCM for fading channel, maximizing the coding gain is a separate problem in itself, as it heavily depends on the trellis and the signal set used. 
\subsection{Code Design Examples}
In all the trellis diagrams shown, the labellings on the edges are shown to the left of each state. The triple inside $[.]$, when read from left to right, denotes the labellings $(\mathcal{X}_{s_1},\mathcal{X}_{r},\mathcal{X}_{s_2})$ corresponding to the edges emerging from top to bottom (the labellings $\mathcal{X}_{s_1},\mathcal{X}_{r},\mathcal{X}_{s_2}$ are shown in the same trellis diagram instead of three different trellis diagrams).
In Examples 2, 3, 4 and 5 considered in this subsection, S transmits two information bits to D in the two phases of relaying. 
 \begin{example}
\begin{figure}[htbp]
\centering
\includegraphics[totalheight=2in,width=2in]{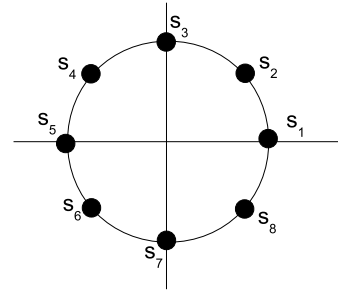}
\caption{8-PSK signal set}	
\label{fig:8psk}	
\end{figure}

 \begin{figure}[htbp]
 \centering
\includegraphics[totalheight=1.5in,width=3.5in]{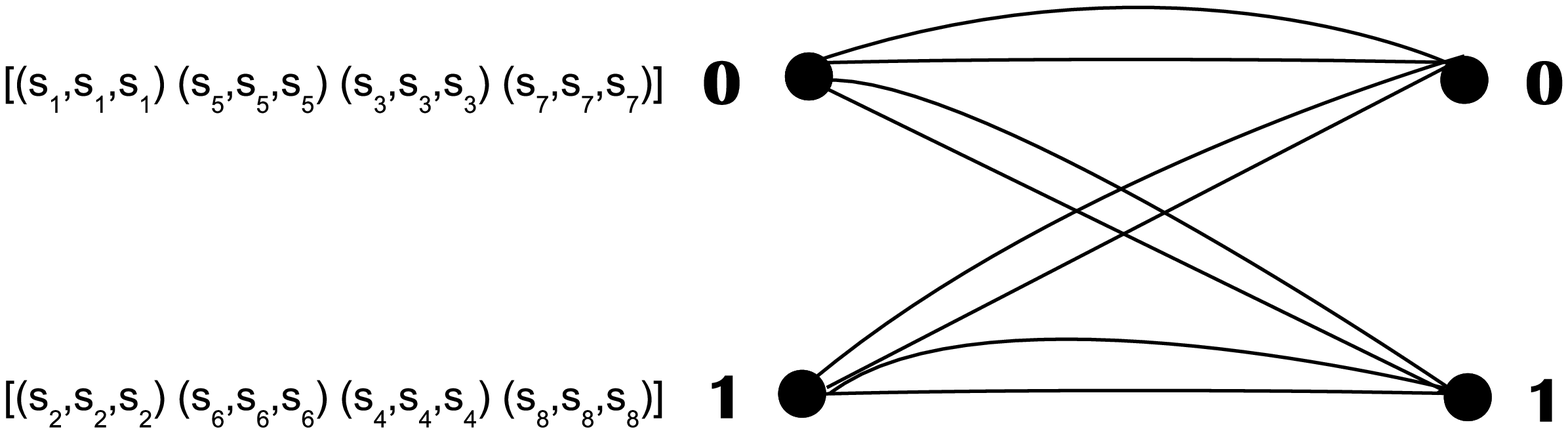}
\caption{Two State Trellis}	
\label{fig:2state}	
\end{figure} 
Consider the case where encoding at S and R take place using a two state trellis (Fig. \ref{fig:2state}) with 8 PSK signal set with signal points labelled as shown in Fig. \ref{fig:8psk}. Let $\delta_0$, $\delta_1$, $\delta_2$ and $\delta_3$ denote the ordered squared Euclidean distances between the points in the 8 PSK signal set, i.e., $\delta_0=(2-\sqrt{2})=0.586$, $\delta_1=2$, $\delta_2=(2+\sqrt{2})=3.414$ and $\delta_3=4$. Since the unmerged length of the code is equal to one, the diversity order cannot exceed two. The choice of $\mathcal{X}_{s_1}$ shown in Fig. \ref{fig:2state} results in the value of the effective length of the trellis $\mathcal{T}_{S_1}$ equal to one and maximizes  $\mathcal{G}_1$. The value of $\mathcal{G}_1=\delta_3=4$. Choosing $\mathcal{X}_{r}$ and $\mathcal{X}_{s_2}$ same as $\mathcal{X}_{s_1}$ results in the value of the effective length of $\mathcal{T}_{S_2}$ equal to one and maximizes  $\mathcal{G}_2$.
 For this example, the diversity order is 2 and the coding gain metric $\mathcal{G}=\delta_3(\delta_3+\gamma\delta_3)=16.5056$, for $\gamma=0.0316$.
 \end{example}
 \begin{example}
 Consider the case where encoding at S and R take place using a four state trellis (Fig. \ref{fig:4state}) with 8 PSK signal set. The unmerged length of the code is two and hence the diversity order cannot exceed four. From \cite{JaLe}, the choice of $\mathcal{X}_{s_1}$ shown in Fig. \ref{fig:4state} results in a value of the effective length of $\mathcal{T}_{S_1}$ equal to two and $\mathcal{G}_1=\delta_0\delta_3=2.344$. 
 
If $\mathcal{X}_{s_1}$, $\mathcal{X}_{s_2}$ and $\mathcal{X}_{r}$ are chosen based on the design criteria for the fading channel (Jamali et al. labelling \cite{JaLe}), the value obtained for the metric $\mathcal{G}_2=(\delta_0\delta_3)^2=5.49$. Instead, if the labellings $\mathcal{X}_{s_1}$, $\mathcal{X}_{s_2}$ and $\mathcal{X}_{r}$ are chosen as shown in Fig. \ref{fig:4state} the value of  $\mathcal{G}_2=\delta_0\delta_1\delta_2\delta_3=16$ and the effective length of $\mathcal{T}_{S_2}$ equal to two.
For this example, the diversity order is 4 and the coding gain metric $\mathcal{G}=\delta_0\delta_3(\delta_1\delta_2+\gamma\delta_0\delta_3)=16.1735$, for $\gamma=0.0316$. 
\begin{figure}[htbp]
\centering
\includegraphics[totalheight=3in,width=3.8in]{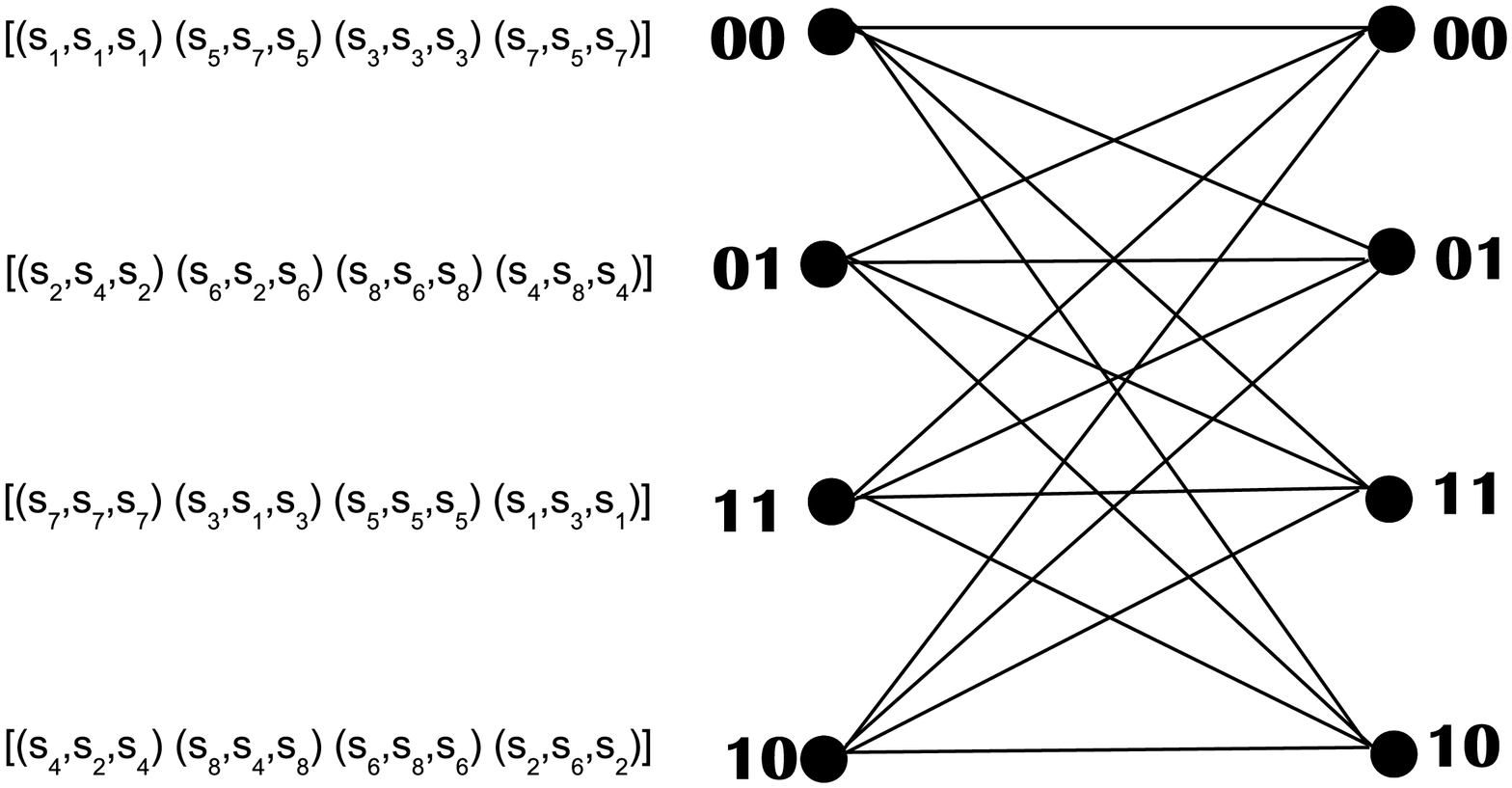}
\vspace{-1 cm}
\caption{Four State Trellis}	
\label{fig:4state}	
\vspace{-0.2 cm}
\end{figure}
\end{example}
 \begin{example}
  Consider the case where encoding at S and R take place using an eight state trellis (Fig. \ref{fig:8state}) with 8 PSK signal set. For this trellis, the diversity order cannot exceed four since the unmerged length of the code is two. From \cite{ScCo}, the choice of $\mathcal{X}_{s_1}$ shown in Fig. \ref{fig:8state} (which is the same as Ungerboeck's labelling \cite{Un}) results in a value of the effective length equal to two and $\mathcal{G}_1=\delta_1\delta_3=8$. Choosing $\mathcal{X}_{r}$ and $\mathcal{X}_{s_2}$ same as $\mathcal{X}_{s_1}$ results in $\mathcal{G}_2=(\delta_1\delta_3)^2=64$ and ensures that the effective length of in $\mathcal{T}_{S_2}$ is equal to two. 
For this example, the diversity order is 4 and the coding gain metric $\mathcal{G}=(\delta_1\delta_3)^2(1+\gamma)=66.0224$, for $\gamma=0.0316$. Increasing the number of states from 4 to 8, while it provides an increase in the coding gain, does not provide diversity advantage. 
\begin{figure}[htbp]
\centering
\includegraphics[totalheight=4in,width=3.75in]{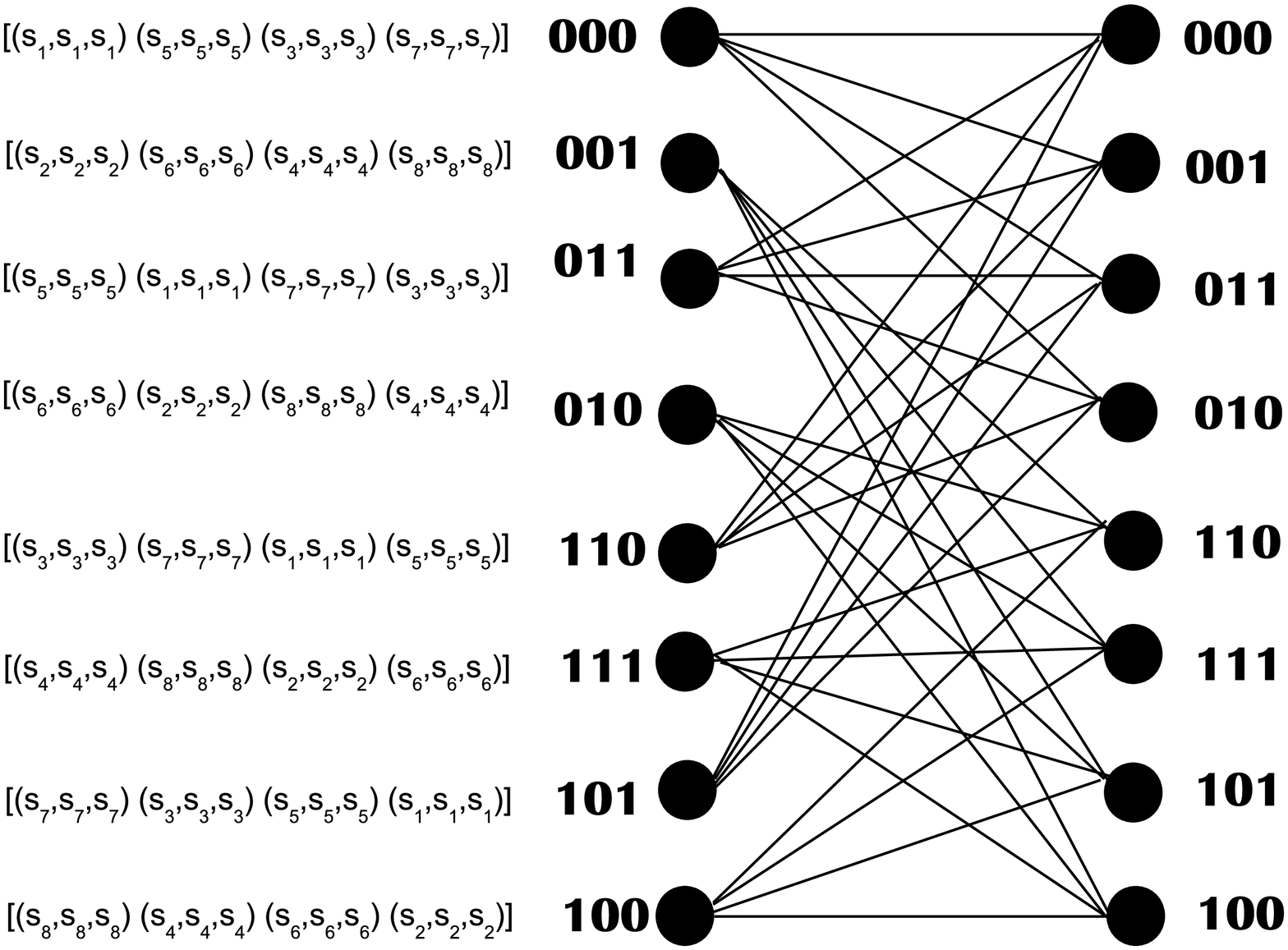}
\vspace{-1 cm}
\caption{8 State Trellis}	
\label{fig:8state}
\vspace{-0.5 cm}	
\end{figure}
 \end{example}
 
 \begin{example}
 Consider the case where encoding at S and R take place using a sixteen state trellis shown in Fig. \ref{fig:16state_8psk}. The unmerged length of the code is three and hence the diversity order cannot exceed six. From \cite{JaLe}, the choice of $\mathcal{X}_{s_1}$ shown in Fig. \ref{fig:16state_8psk} (which is the same as Ungerboeck's labelling \cite{Un}) results in a value of the effective length of $\mathcal{T}_{S_1}$ equal to three and $\mathcal{G}_1=\delta_0\delta_1\delta3=4.68$. 
 Labellings $\mathcal{X}_{r}$ and $\mathcal{X}_{s_2}$ chosen to be the same as $\mathcal{X}_{s_1}$ result in  $\mathcal{G}_2=(\delta_0\delta_1\delta_3)^2=21.96$ and the effective length of $\mathcal{T}_{S_2}$ equal to three.
For this example, the diversity order is 6 and the coding gain metric $\mathcal{G}=(\delta_0\delta_1\delta_3)^2(1+\gamma)=22.66$, for $\gamma=0.0316$. 
\begin{figure}[htbp]
\centering
\includegraphics[totalheight=6in,width=3.75in]{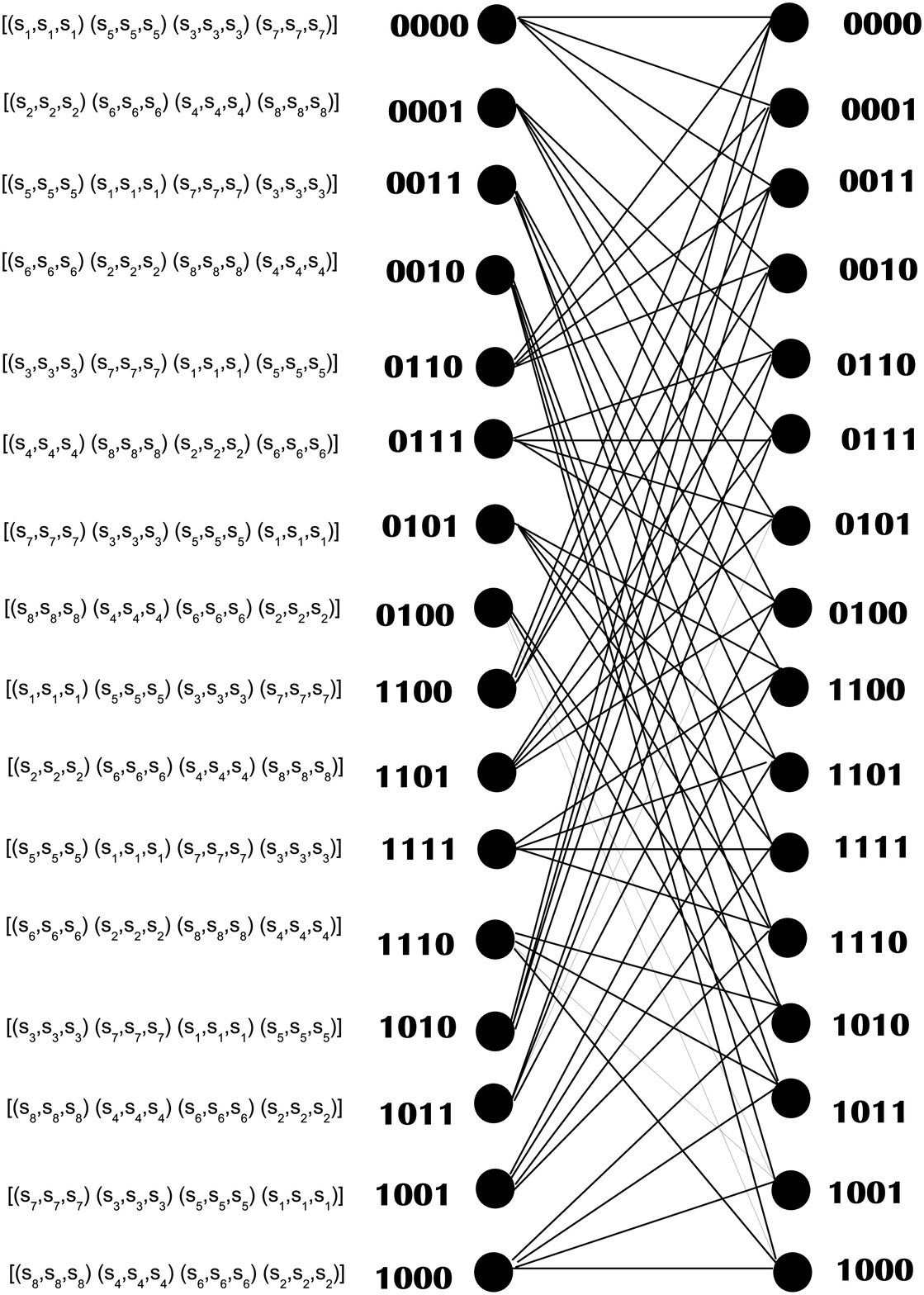}
\vspace{-1 cm}
\caption{16 State Trellis}	
\label{fig:16state_8psk}
\vspace{-0.5 cm}	
\end{figure}
\end{example}
 In the following example we consider the case where 3 bits of information get transmitted from S to D in the two phases of relaying.
\begin{example}
 \begin{figure}[htbp]
\centering
\includegraphics[totalheight=2in,width=2in]{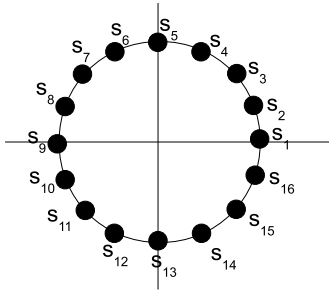}
\vspace{-0.5 cm}
\caption{16 PSK Signal Set}	
\label{fig:16psk}	
\end{figure} 
  Consider the case where encoding at S and R take place using an eight state trellis (Fig. \ref{fig:8state_16psk}) with 16 PSK signal set whose points are labelled as shown in Fig. \ref{fig:16psk}. For this trellis the diversity order cannot exceed four, since the unmerged length of the code is two. The choice of $\mathcal{X}_{s_1}$ shown in Fig. \ref{fig:8state_16psk} results in a value of $\mathcal{G}_1=  0.0892$ and ensures that the effective length of $\mathcal{T}_{S_1}$ is two.
  The labellings $\mathcal{X}_{r}$ and  $\mathcal{X}_{s_2}$ shown in Fig. \ref{fig:8state_16psk} result in $\mathcal{G}_2=0.1177$ and ensures that the effective length of $\mathcal{T}_{S_2}$ is two.
 For this example, the diversity order is 4 and the coding gain metric $\mathcal{G}=0.1295$, for $\gamma=0.1$. 
\begin{figure*}[htbp]
\centering
\includegraphics[totalheight=4in,width=8in]{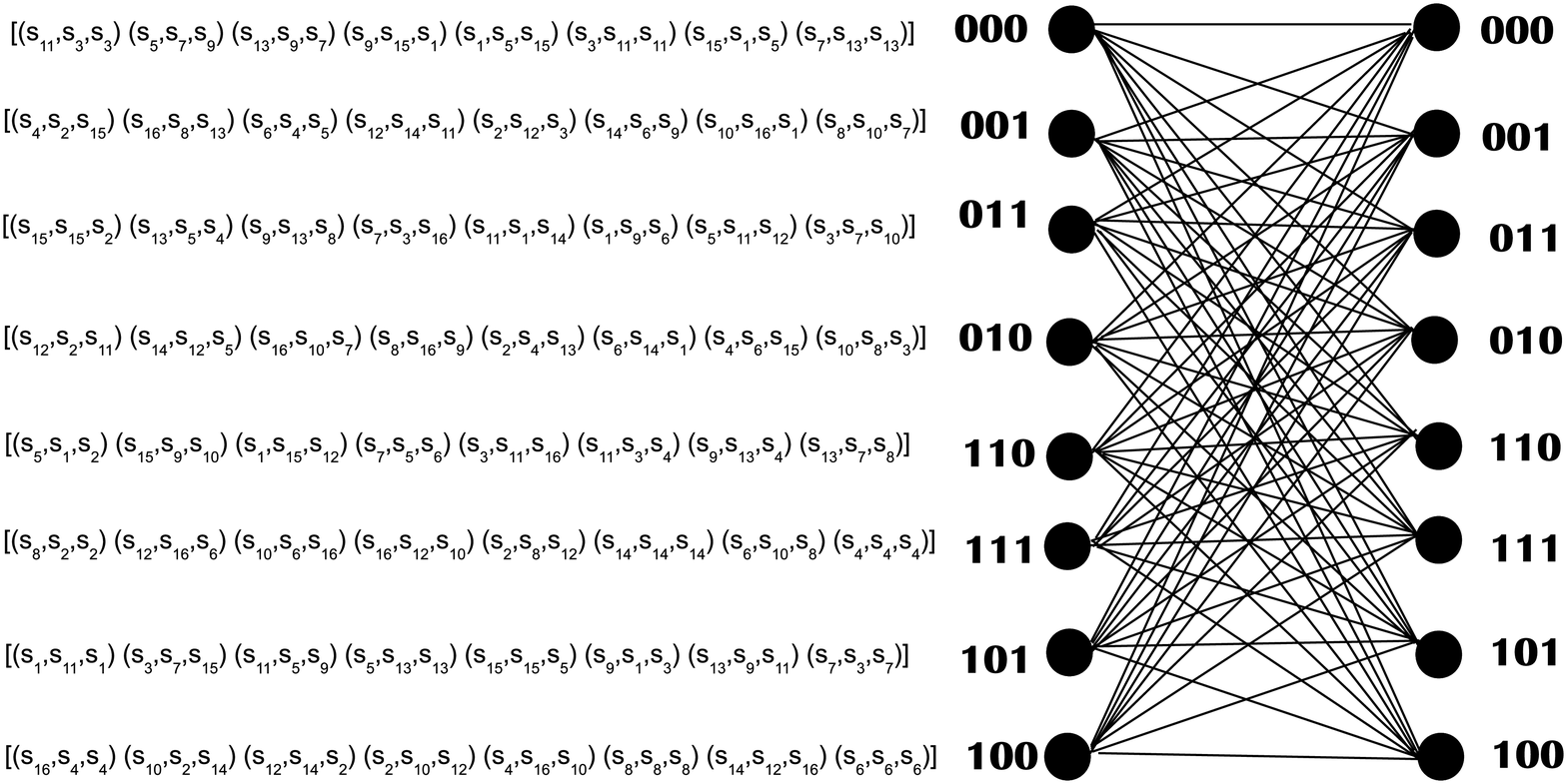}
\vspace{-1 cm}
\caption{8 State Trellis}	
\label{fig:8state_16psk}	
\end{figure*}
 \end{example}
 \section{SIMULATION RESULTS}
%
\begin{table*}
\setlength{\tabcolsep}{1pt}
\caption{Comparison of uncoded 8 PSK transmission scheme and 16 PSK TCM Schemes}
 \centering
 \label{table9}
 \begin{tabular}{|c|c|c|c|}
\hline Scheme & Diversity Order &  $E_\mathcal{S}$ in dB for $BER=10^{-4}$\\ 
\hline Uncoded Transmission Scheme - 8 PSK - With Constant Labelling       & 2     & 20 dB \\ 
\hline Uncoded Transmission Scheme - 8 PSK - With Our Labelling            & 2     & 18 dB \\ 
\hline Relay Channel - 8 State TCM                                         & 4     & 12.5 dB \\ 
\hline 
\end{tabular} 
\end{table*}
\begin{figure}[htbp]
\centering
\includegraphics[totalheight=3in,width=3.75in]{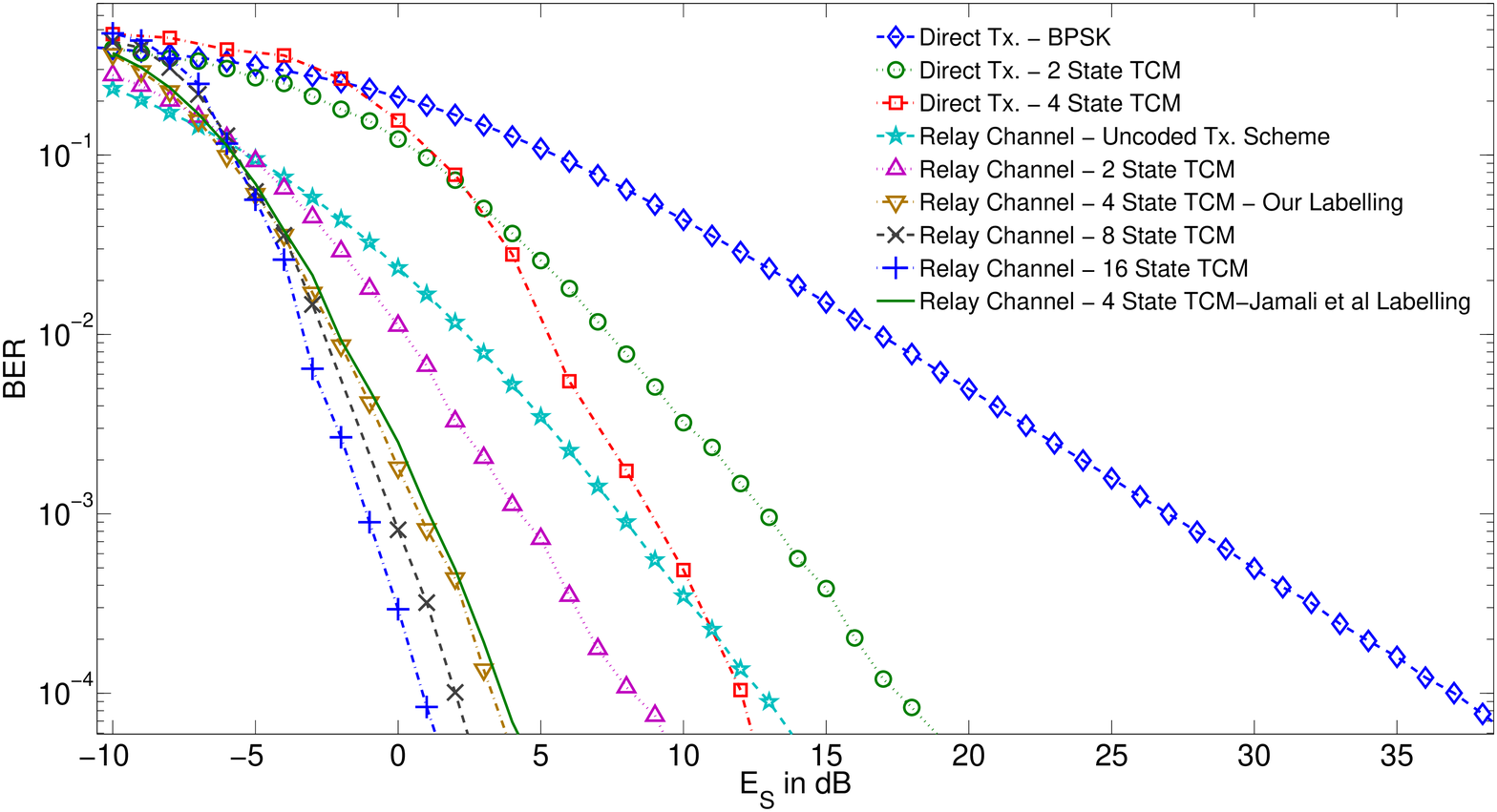}
\caption{$E_\mathcal{S}$ vs $BER$ for the uncoded transmission scheme using 4 PSK and TCM Schemes for $\sigma^2_{sd}=0$ dB, $\sigma^2_{sr}=10$ dB and $\sigma^2_{rd}=10$ dB}	
\label{fig:ber_comp}	
\end{figure}
\begin{figure}[htbp]
\centering
\includegraphics[totalheight=3in,width=3.75in]{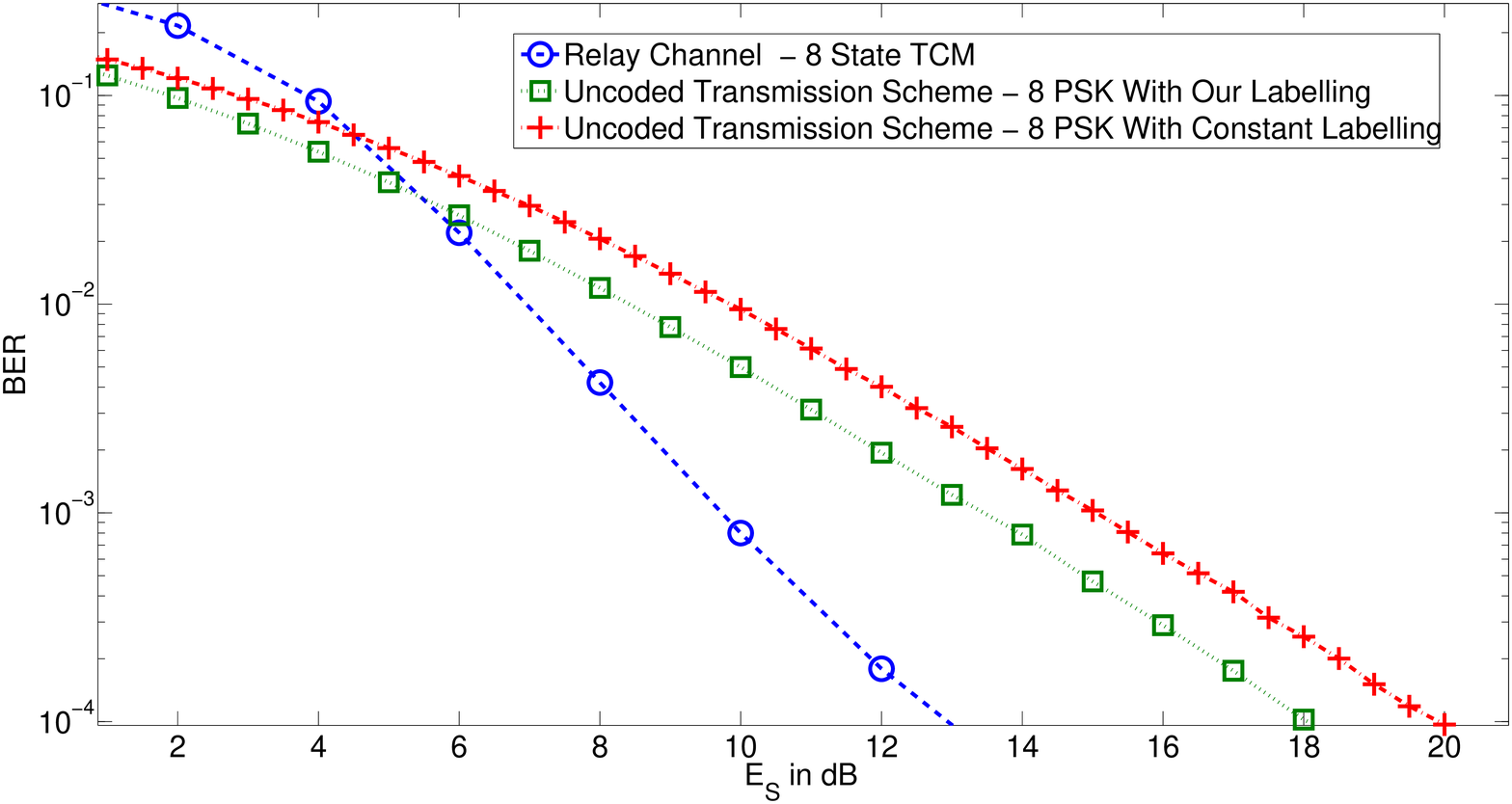}
\caption{$E_\mathcal{S}$ vs $BER$ for the uncoded transmission scheme using 8 PSK and TCM Schemes for $\sigma^2_{sd}=0$ dB, $\sigma^2_{sr}=10$ dB and $\sigma^2_{rd}=10$ dB}	
\label{fig:ber_comp_16psk}	
\end{figure}


\begin{figure}[htbp]
\centering
\includegraphics[totalheight=3in,width=3.75in]{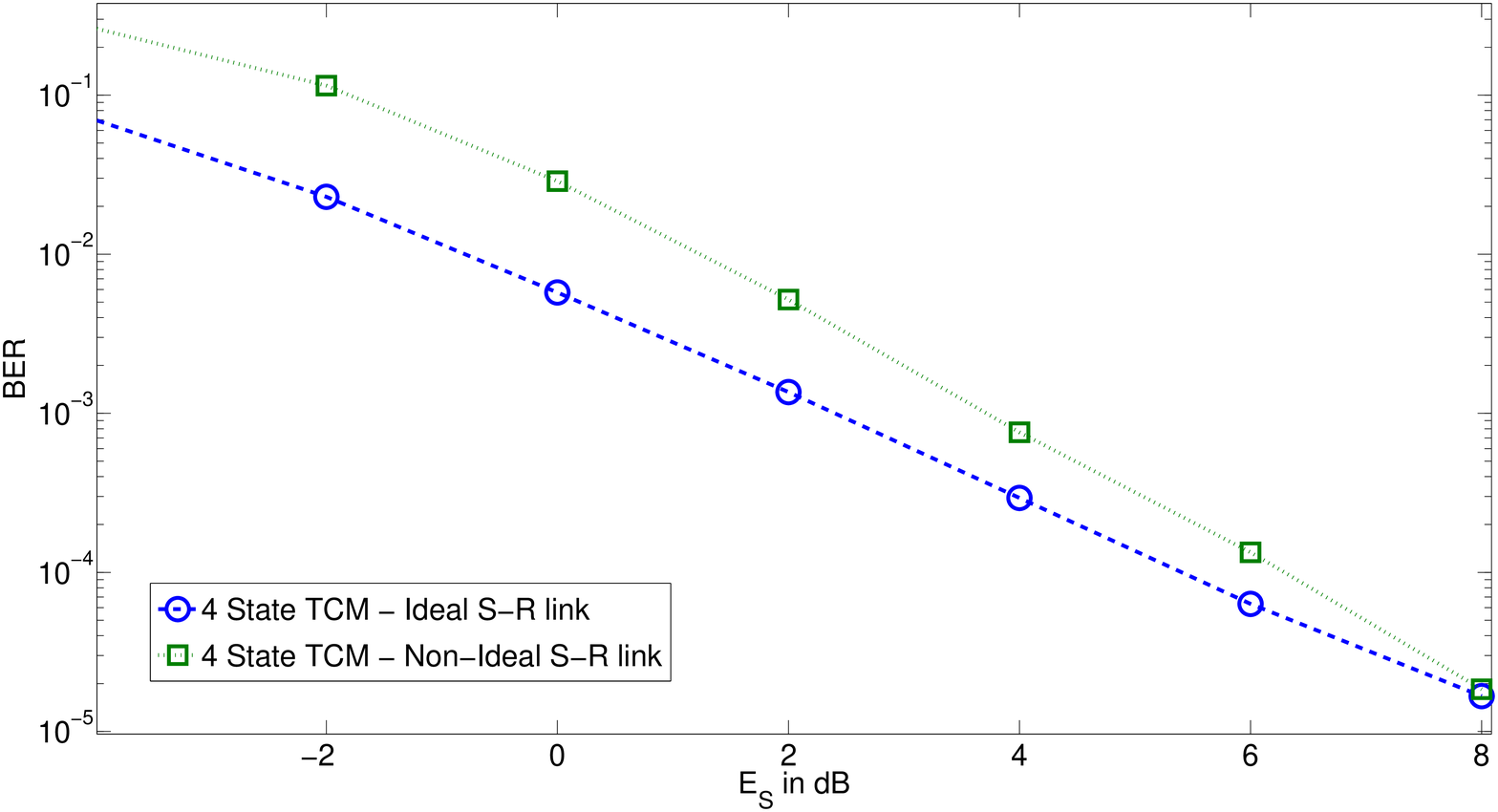}
\caption{$E_\mathcal{S}$ vs $BER$ comparison for the proposed TCM scheme with ideal and non-ideal S-R links for $\sigma^2_{sd}=0$ dB, $\sigma^2_{sr}=10$ dB and $\sigma^2_{rd}=10$ dB}	
\label{fig:ideal_nonideal_comp}	
\end{figure}


 The $E_\mathcal{S}$ vs $BER$ plots for the different schemes achieving a spectral efficiency of 1 bpcu are shown in Fig. \ref{fig:ber_comp}, for $\sigma^2_{sd}=0$ dB, $\sigma^2_{sr}=15$ dB and $\sigma^2_{rd}=15$ dB. 
The diversity order and the value of $E_\mathcal{S}$ required at a BER of $10^{-4}$ are summarized in Table \ref{table5}. 
 As observed in the previous section, the uncoded transmission scheme provides a diversity order 2. We observe that using a 2 state trellis provides a diversity order 2 as expected. At a BER of $10^{-4}$, the 2 state 8 PSK TCM scheme provides 4.5 dB gain over the uncoded transmission scheme using 4 PSK. When the number of states is increased to 4, we see that the diversity order increases to 4 as predicted and at a BER of $10^{-4}$ a large gain of 9.5 dB is obtained over the uncoded transmission scheme using 4 PSK. Also, it can be seen from Fig. \ref{fig:ber_comp}, it can be seen that the 4 state 8 PSK scheme with our labelling provides a gain of 0.6 dB over the case when S and R use Jamali et al. labelling. Increasing the number of states to 8, from Fig. \ref{fig:ber_comp} the diversity order stays at 4 but a coding gain of 1.5 dB is obtained over the 4 state 8 PSK TCM scheme.  With the 16 state 8 PSK TCM scheme, a diversity order of 6 is obtained and a gain of 1.5 dB is obtained over the 8 state 8 PSK TCM scheme. In essence, at a BER of $10^{-4}$, a large gain of 12.5 dB is obtained  using the 16 state 8 PSK TCM scheme over the uncoded transmission scheme using 4 PSK.  
 
 Fig. \ref{fig:ber_comp_16psk} shows the plots comparing the uncoded transmission scheme using 8 PSK with our labelling and with constant labelling, and the 8 state 16 PSK TCM scheme, for $\sigma^2_{sd}=0$ dB, $\sigma^2_{sr}=10$ dB and $\sigma^2_{rd}=10$ dB. At a BER of $10^{-4}$ the uncoded transmission scheme using 8 PSK with our labelling and 8 state 16 PSK TCM  provide a gain of 2 dB and 7.5 dB respectively over the 8 PSK uncoded transmission scheme with constant labelling. The diversity order and the value of $E_\mathcal{S}$ required at a BER of $10^{-4}$ are summarized in Table \ref{table9}.
 
 
 Fig. \ref{fig:ideal_nonideal_comp} shows the $E_\mathcal{S}$ vs $BER$ plot for the 4 state 8 PSK TCM scheme, for the cases where the S-R link is ideal and non-ideal, for $\sigma^2_{sd}=0$ dB, $\sigma^2_{sr}=10$ dB and $\sigma^2_{rd}=10$ dB. As observed in Section IV E, the high SNR performance of the proposed near-ML decoder with non-ideal S-R link approaches the performance of the optimal ML decoder with an ideal S-R link. 

\section{DISCUSSION}
A TCM scheme for the half duplex fading relay channel was proposed. A near-ML decoder whose high SNR performance approaches the performance of the optimal ML decoder was obtained. Based on the expression for PEP bounds, code design criteria to maximize the diversity order and coding gain were formulated. Maximizing the diversity order can be done by a proper choice of the labellings used at the source during Phase 1 and Phase 2. A unified procedure to find the labelling scheme which maximizes the coding gain for all signal sets and trellises is not known and will be an interesting topic for further research. 
\section*{Acknowledgement}
This work was supported  partly by the DRDO-IISc program on Advanced Research in Mathematical Engineering through a research grant as well as the INAE Chair Professorship grant to B.~S.~Rajan.
 
\end{document}